\documentclass[12pt]{article}

\usepackage{latexsym}

\font\tenmsbm=msbm10 scaled 1200
\font\sevenmsbm=msbm9
\newfam\msbmfam
\textfont\msbmfam=\tenmsbm
\scriptfont\msbmfam=\sevenmsbm

\makeatletter
\@addtoreset{equation}{section}
\makeatother

\newcommand{\eref}[1]{(\ref{#1})}

\def\beq{\begin{equation}}
\def\eeq{\end{equation}}
\def\bea{\begin{eqnarray}}
\def\eea{\end{eqnarray}}
\def\bet{\begin{tabular}}
\def\eet{\end{tabular}}

\def\ve{\varepsilon}

\def\no{\nonumber}

\catcode`@=11
\def\lsim{\mathchoice
  {\mathrel{\lower.8ex\hbox{$\displaystyle\buildrel<\over\sim$}}}
  {\mathrel{\lower.8ex\hbox{$\textstyle\buildrel<\over\sim$}}}
  {\mathrel{\lower.8ex\hbox{$\scriptstyle\buildrel<\over\sim$}}}
  {\mathrel{\lower.8ex\hbox{$\scriptscriptstyle\buildrel<\over\sim$}}} }
\def\gsim{\mathchoice
  {\mathrel{\lower.8ex\hbox{$\displaystyle\buildrel>\over\sim$}}}
  {\mathrel{\lower.8ex\hbox{$\textstyle\buildrel>\over\sim$}}}
  {\mathrel{\lower.8ex\hbox{$\scriptstyle\buildrel>\over\sim$}}}
  {\mathrel{\lower.8ex\hbox{$\scriptscriptstyle\buildrel>\over\sim$}}} }
\def\croce{\displaystyle / \kern-0.2truecm\hbox{$\backslash$}}
\def\lqua{\lower4pt\hbox{\kern5pt\hbox{$\sim$}}\raise1pt
\hbox{\kern-8pt\hbox{$<$}}~}
\def\gqua{\lower4pt\hbox{\kern5pt\hbox{$\sim$}}\raise1pt
\hbox{\kern-8pt\hbox{$>$}}~}
\def\mma{\lower1pt\hbox{\kern5pt\hbox{$\scriptstyle <$}}\raise2pt
\hbox{\kern-7pt\hbox{$\scriptstyle >$}}~}
\def\mmb{\lower1pt\hbox{\kern5pt\hbox{$\scriptstyle >$}}\raise2pt
\hbox{\kern-7pt\hbox{$\scriptstyle <$}}~}
\def\mmc{\lower4pt\hbox{\kern5pt\hbox{$<$}}\raise1pt
\hbox{\kern-8pt\hbox{$>$}}~}
\def\mmd{\lower4pt\hbox{\kern5pt\hbox{$>$}}\raise1pt
\hbox{\kern-8pt\hbox{$<$}}~}
\def\lsu{\raise4pt\hbox{\kern5pt\hbox{$\sim$}}\lower1pt
\hbox{\kern-8pt\hbox{$<$}}~}
\def\gsu{\raise4pt\hbox{\kern5pt\hbox{$\sim$}}\lower1pt
\hbox{\kern-8pt\hbox{$>$}}~}
\def\croce{\displaystyle / \kern-0.2truecm\hbox{$\backslash$}}
\def\ali{\hbox{A \kern-.9em\raise1.7ex\hbox{$\scriptstyle \circ$}}}
\def\2frecce{\hbox{\lower 0.3ex\hbox{$\leftarrow$} 
\hbox{\kern-1.3em\raise 0.3ex\hbox{$\rightarrow$}}}}
\def\quad@rato#1#2{{\vcenter{\vbox{
        \hrule height#2pt
        \hbox{\vrule width#2pt height#1pt \kern#1pt \vrule width#2pt}
        \hrule height#2pt} }}}
\def\quadratello{\mathchoice
\quad@rato5{.5}\quad@rato5{.5}\quad@rato{3.5}{.35}\quad@rato{2.5}{.25} }

\def\Ai{\hbox{\hbox{${\cal A}$}}\kern-1.9mm{\hbox{${/}$}}}
\def\Vi{\hbox{\hbox{${\cal V}$}}\kern-1.9mm{\hbox{${/}$}}}
\def\Di{\hbox{\hbox{${\cal D}$}}\kern-1.9mm{\hbox{${/}$}}}
\def\lam{\hbox{\hbox{${\lambda}$}}\kern-1.6mm{\hbox{${/}$}}}
\def\D{\hbox{\hbox{${D}$}}\kern-1.9mm{\hbox{${/}$}}}
\def\A{\hbox{\hbox{${A}$}}\kern-1.8mm{\hbox{${/}$}}}
\def\V{\hbox{\hbox{${V}$}}\kern-1.9mm{\hbox{${/}$}}}
\def\parz{\hbox{\hbox{${\partial}$}}\kern-1.7mm{\hbox{${/}$}}}
\def\B{\hbox{\hbox{${B}$}}\kern-1.7mm{\hbox{${/}$}}}
\def\R{\hbox{\hbox{${R}$}}\kern-1.7mm{\hbox{${/}$}}}
\def\si{\hbox{\hbox{${\xi}$}}\kern-1.7mm{\hbox{${/}$}}}

\begin{document}
\begin{titlepage}
\begin{flushright}
Preprint DFPD 00/TH/33\\
\end{flushright}
\vspace{1truecm}
\begin{center}

{\Large \bf Spin--statistics transmutation}

\par

\vspace{0.4cm}
{\Large \bf in relativistic quantum field theories of dyons}

\vspace{1cm}
{K. Lechner\footnote{kurt.lechner@pd.infn.it} and
{P.A. Marchetti}\footnote{pieralberto.marchetti@pd.infn.it}}

\vspace{1cm}
{\it Dipartimento di Fisica, Universit\`a  degli Studi di Padova,

Istituto Nazionale di Fisica Nucleare, Sezione di Padova, 

Via F. Marzolo 8, 35131 Padova, Italia}
\vskip 1truecm

\begin{abstract}
\vskip 0.5truecm
We analyse spin and  statistics of quantum dyon fields, i.e. fields 
carrying both electric and magnetic charge, in 3+1 space--time dimensions.
It has been shown long time ago that, at the quantum mechanical level,
a composite dyon made out of a magnetic pole
of charge $g$ and a particle of electric charge $e$ possesses half--integral
spin and fermionic statistics, if the constituents are bosons and the Dirac
quantization condition $eg=2\pi n$ holds, with $n$ {\it odd}. 
This  phenomenon is called  
spin--statistics transmutation. We show that the same phenomenon occurs at 
the quantum field theory level for an {\it elementary} dyon.
This analysis requires the construction of gauge invariant charged dyon 
fields. Dirac's proposal for such fields, relying on a Coulomb--like photon
cloud, leads to quantum correlators exhibiting an unphysical 
dependence on the Dirac--string. Recently Froehlich and Marchetti proposed 
a recipe for charged dyon fields, based on a sum over Mandelstam--strings,
which overcomes this problem. Using this recipe we derive explicit 
expressions for the quantum field theory correlators and we provide a  
proof of the occurrence of spin--statistics transmutation. The proof 
reduces to a computation of the
self--linking numbers of dyon worldlines
and Mandelstam strings, projected on a fixed time three--space. 
Dyon composites are also analysed.
The transmutation discussed in this paper bares some analogy with the 
appearance of 
anomalous spin and statistics for particles or vortices in Chern--Simons 
theories in 2+1 dimensions. However, peculiar features appear in 3+1 
dimensions e.g. in the spin addition rule. 

\end{abstract}

\end{center}
\end{titlepage}

\newpage

\baselineskip 6 mm

\section{Introduction}

In this paper we   analyse  spin and statistics of 
quantum dyon fields, i.e. fields carrying both electric and magnetic 
charge in 3+1 space--time dimensions, relying upon the construction of 
their correlation functions sketched in \cite{1}.

Previous discussions about this subject were all based either on a quantum 
mechanical analysis of {\it composite} dyons, i.e. composites of a magnetic 
monopole and a charged particle \cite{2,3}, or on a semiclassical 
treatment of quantum
field theories where dyons appear \cite{4}, but never, at least to our 
knowledge, on an analysis of ``elementary dyons" in a fully quantized 
field theory. With ``elementary dyon" we mean a point--like particle
which carries electric as well as magnetic charge.

Both at the classical and quantum mechanical levels one finds indications that 
composite dyons may exhibit anomalous spin.
Classically one can compute the angular momentum $\vec J$ stored in 
the electromagnetic field generated by a magnetic pole with magnetic charge
$g$ located at the origin and an electric charge of strength $e$ at a 
distance $a$ along the 3--axis. For the electric field we have 
$\vec E={e\over 4\pi}{\vec x-\vec a\over |\vec x-\vec a|^3}$ and for the 
magnetic field we have $\vec B={g\over 4\pi}{\vec x\over |\vec x|^3}$.
The only non--vanishing component of the angular momentum is then
\beq
J_3 = \int d^3 x [\vec x \wedge( \vec E \wedge \vec B)]_3
={eg \over 4\pi},
\label{1.1}
\eeq
where the last equality holds for $a \not= 0$. 
The quantum mechanical requirement that the total angular momentum is 
quantized with spectrum contained in ${{\bf Z}\over 2}$ $(\hbar = 1)$ 
reproduces the Dirac quantization condition for the coexistence of 
electrically and magnetically charged particles:
\beq\label{1.2}
eg \in 2\pi {\bf Z}. 
\eeq
In particular if ${eg\over 2\pi}$ is odd, the classical calculation 
suggests that the composite dyon carries half--integral spin.

These ideas  can be made mathematically precise \cite{5} and it has been 
proved that
the Hilbert space of states of the composite dyon carries a projective 
representation of the rotation group, provided Dirac's quantization condition 
\eref{1.2} holds, and under this condition the wave function acquires a phase 
factor $e^{i {eg \over 2}}$ under a $2\pi$--rotation.

Subsequently it has been shown that for such dyons the usual 
spin--statistics connection holds, so that if the constituent particles 
are bosons and $eg = 2\pi$ the composite behaves as a fermion with 
half--integral spin \cite{3,6}. We call this phenomenon ``spin--statistics 
transmutation" borrowing a terminology frequently used in 2+1 dimensional 
systems \cite{7}.

Let us turn to ``elementary" point--like dyons. If we set $a=0$ in the 
calculation of the classical angular momentum, we find that the integral 
in \eref{1.1}  vanishes. Hence it is far from clear what happens at 
the quantum level for elementary dyons. One of the main purposes of this 
paper is to clarify this issue in the framework of relativistic quantum 
field theory.

Several problems have to be solved along the way to define a consistent 
setting. The classical equations of motion of a relativistic 
point--like dyon (coupled to the electromagnetic field) are Lorentz covariant, 
reading as

\beq\label{1.3}
\partial^\mu F_{\mu\nu} (x) = e \int d\tau {dy_\nu (\tau)\over d\tau} 
\delta^{(4)} (x - y (\tau))
$$
$$
\partial^\mu * F_{\mu\nu} (x) = g \int d\tau {dy_\nu (\tau)\over d\tau} 
\delta^{(4)} (x-y (\tau))
$$
$$
m {d^2 y^\mu\over d\tau^2} = (e F^{\mu\nu}  + g *F^{\mu\nu}) {dy_\nu\over 
d\tau},
\eeq
where $y^\mu (\tau)$ parametrizes the particle trajectory,
$\tau$ is the proper time, 
$*F_{\mu\nu} = {1\over 2} \varepsilon_{\mu\nu\rho\sigma} F^{\rho\sigma}$
and $m$ is the mass of the dyon.
However, the implementation of a Lorentz--invariant action principle,
even for classical point--like particles, encounters some difficulty.
The main reason for this is that an action principle for Maxwell's 
equations requires necessarily the introduction of vector potentials.
Eventually the action for classical 
point--like particles 
turns out to be Lorentz--invariant modulo some integer, if Dirac's 
condition holds.

The situation becomes even worse at the level of the quantum field
theory. In a functional integral approach the fundamental 
ingredients are a classical {\it field} theory action and the corresponding
classical equations of motion for the dyon fields. Contrary to what happens 
for the 
system \eref{1.3} for point--like particles, the equations for dyon fields,
e.g. Dirac's equations,  
involve the vector potential explicitly through a covariant derivative.
But this potential is a priori not 
well defined because the presence of monopole--like configurations 
introduces unphysical Dirac--strings, which are in a sense 
``global" gauge artifacts.

As a matter of fact, a consistent classical field theory of dyons does not 
exist and the possibility of a consistent setting for quantum fields relies
upon a realization of field--particle quantum duality \cite{8}, which 
allows, roughly speaking, to write the
correlation functions of local (with compact support) gauge--invariant 
observables in terms of a consistent quantum mechanics of point--like dyons.
As shown in \cite{8,iengo} these correlation functions are independent 
of the position of Dirac--strings provided some
form of ``Dirac quantization condition" is imposed.

It turns out that there are two inequivalent consistent classes of 
quantum field theories of 
dyons, characterized by their (S--) duality groups \cite{1}. 
At the classical level one
 verifies easily that the equations of motion \eref{1.3} are invariant under an 
$SO(2)$ group of transformations, parametrized by an angle $\theta$, defined 
by
\beq\label{4}
\left(\matrix{F^{\mu\nu} \cr
*F^{\mu\nu}\cr}\right)\Longrightarrow \left(\matrix{\cos\theta & \sin\theta 
\cr
-\sin\theta & \cos\theta\cr}\right) \left(\matrix{F^{\mu\nu} \cr
*F^{\mu\nu}\cr} \right)
$$
$$
\left(\matrix{e\cr
g\cr} \right) \Longrightarrow \left(\matrix{\cos\theta &  \sin\theta \cr
-\sin\theta & \cos\theta\cr} \right) \left(\matrix{e \cr
g\cr} \right). 
\eeq
Let us consider a quantum field theory of dyons of $N$ species with charges 
$\{e_r, g_r\}$, $r=1, ..., N$. One can show that there is a 
consistent quantum field 
theory preserving only the ${\bf Z}_4$ subgroup of $SO(2)$ generated by 
the rotation of an angle $\theta = {\pi\over 2}$ (common to all species), 
provided the Dirac quantization condition
\beq\label{5}
e_r g_s \in 2\pi {\bf Z} 
\eeq
holds.

There is also a consistent quantum field theory preserving the full 
$SO(2)$ duality
group, provided the ``Schwinger--Zwanziger"  quantization 
condition
\beq\label{6}
{1\over 2} (e_r g_s - e_s g_r) \in 2\pi {\bf Z} 
\eeq
holds.

Actually, from the correlation functions of local fields mentioned above one
can only reconstruct the neutral observables and the vacuum sector of the 
quantum field theory, i.e. the physical states which are electrically and 
magnetically neutral; for the construction of charged field operators and 
of charged sectors one needs something more.

Owing to a theorem discussed by Strocchi \cite{9}, gauge--invariant charged 
fields must be necessarily non--local. A natural prescription for the 
construction of 
electrically charged fields was given by Dirac \cite{10}, and it corresponds
essentially 
to a gauge--invariant dressing of the non gauge--invariant local
charged field (e.g. a scalar $\hat\phi$ of charge $e$ in scalar 
QED), by multiplying it with an 
exponential of the photon field $\hat A$, weighted by a classical Coulomb 
field $\vec E$, generated by a unit charge located at $\vec x$:
\beq
\hat\phi (\vec x) \Longrightarrow  \hat\phi (\vec x) \,e^{i e\int
\hat A (\vec y) \cdot \vec E (\vec x - \vec y) d^3 y}.
\eeq
However, this procedure becomes inconsistent when electric and magnetic 
dynamical charges coexist. Indeed, one can show that the correlation 
functions of the fields dressed in this way depend on the position of the 
Dirac--strings, even if some quantization condition for the charges is
imposed.
A proposal for a modification of Dirac's recipe which leads to 
Dirac--string independent correlation functions
has been presented in \cite{11}, 
and in \cite{1} we sketched how this proposal can be adapted to the setting 
of a quantum field theory of dyons. 

Using the charged fields constructed accordingly we will perform in the 
present paper a derivation of spin and statistics of elementary quantum 
dyons with the following results: 1) for a ${\bf
Z}_4$--dyon with charges $e_r$ and $g_r$ such that $e_r g_r/ 2\pi$ is 
{\it odd} spin--statistics 
transmutation from a boson to a fermion (and viceversa) takes place while,
if $e_r g_r/ 2\pi$ is {\it even}, there is no transmutation;
2) for an $SO(2)$--dyon spin and statistics are always the naive ones.

The derivation of these results relies on an explicit representation 
of euclidean correlation functions of charged fields, in terms of a 
sum over currents with support on families of loops (closed particles
trajectories); both spin and statistics
are then related to the self--linking numbers of the projections of these 
loops on a three--dimensional space at fixed time.

In a certain sense, the regularization needed for the definition of the 
self--linking numbers plays the role of the finite distance $a$ between 
the magnetic and electric charge in the classical 
calculation of the angular momentum for
composite dyons, see \eref{1.1}. The results we obtain for 
${\bf Z}_4$--dyons are, in fact,
in agreement with the suggestions obtained from those calculations.

For concreteness we will assume throughout this paper that our dyons carry 
``intrinsic" bosonic statistics, i.e. that they are described by
complex scalar fields (with an appropriate soft photon cloud). This means
that we are analysing the transmutation from bosons to fermions. The 
opposite case, where the dyons are described by complex spinor fields, 
can be analysed with the same techniques \cite{8,kur}
and leads to the same results.

Finally, let us remark that in this paper we use the formalism of 
differential forms  and that we adopt a functional integral approach to 
quantum field theory at a formal level, i.e. ignoring ultraviolet 
divergences. Some U.V. regulator is implicitly understood (e.g. the 
lattice) and we do not discuss its removal; however, we expect our final 
results to be stable under renormalization because spin and statistics 
are large--scale properties.

\section{Relativistic quantum field theory of dyons}

\subsection{Classical point--like particles}

As discussed in the introduction there are two (in general) inequivalent 
classes of relativistic quantum field theories (QFT) of dyons classified
by their duality groups, ${\bf Z}_4$ or $SO(2)$.

Both classes admit different, but equivalent, formulations: \`a la Schwinger 
\cite{12} with one gauge 
potential (with a single string or with two half--weighted antisymmetric 
strings, 
respectively), \`a la Zwanziger \cite{13} 
(with an $i \varepsilon$--prescription 
to invert $n_\mu\partial^\mu$ in the gauge propagator or with the principal 
value prescription, resp.), \`a la PST \cite{14} (with or without an additional 
interaction term ``${1\over 2} C_1 \wedge C_2"$ between Dirac strings, resp.) 
or its dual formulation \cite{15}.

Although the PST formulation has some advantages in that it is manifestly 
Lorentz invariant, in this paper we adopt the formulation \`a 
la Schwinger for the ${\bf Z}_4$--dyons, which are our main  concern 
since they exhibit spin--statistic transmutation; on the other hand, our 
remarks on the 
$SO(2)$--dyons are based on a formulation \`a la Zwanziger.
These formulations are somewhat simpler at the QFT level and this motivates 
our choice.

We start by 
introducing some basic notions in the theory of (de--Rham) currents \cite{16} 
which will be of key importance
in the following.
A $p$--current in ${\bf R}^4$ is a linear functional on the space of 
smooth (4--$p$)--forms with compact support, which is continuous in the 
sense of distributions; i.e. $p$--currents are ``$p$--forms" with 
distribution--valued components.

In the space of currents one can define a map, here denoted by PD, 
extending Poincar\`e duality, which associates to every $p$--dimensional 
surface $\Sigma_p$ a (4--$p$) current $\Phi_{\Sigma_p}$ according to
\bea\nonumber
{\rm PD}: \Sigma_p &\rightarrow& \Phi_{\Sigma_p} \equiv {\rm PD} (\Sigma_p)
\\
\int_{\Sigma_p} \alpha_p &=& \int_{{\bf R}^4} \alpha_p \wedge 
\Phi_{\Sigma_p},
\label{2.1}
\eea
for any smooth $p$--form $\alpha_p$ of compact support.
In particular, the image by PD of a closed surface is a closed current i.e.
if $\partial \Sigma_p = \emptyset$ then $d\Phi_{\Sigma_p} =0$, 
where $\partial$ 
denotes the boundary operator. Via regularization
one can define also the integral of a current 
$\Phi_{\Sigma_p}$ over a generic $4-p$ dimensional surface $S_{4-p}$, and 
the result is
an integer if the integral is well defined and it counts the number of 
intersection with sign, $m$, of $\Sigma_p$ and $S_{4-p}$. We can then formally 
write
\beq\label{2.2}
\int_{{\bf R}^4} \Phi_{\Sigma_p} \wedge \Phi_{S_{4-p}}=m. 
\eeq
Linear combinations of such $p$--currents with integer coefficients are 
called integer currents; they are PD--dual to integer 
(4--$p$)--chains \cite{16}.

After this digression we turn to dyons. Let $\{\gamma_r \equiv y^\nu_r 
(\tau_r)\}$ denote a set of boundaryless worldlines of dyons with 
charges $\{e_r, g_r\}$, 
parametrized by proper times $\tau_r,$.
To such trajectories one can associate 3--currents 
\beq\label{2.3}
J_{(r)} (x) = {1\over 3!} dx^\mu \wedge dx^\nu \wedge dx^\rho
\varepsilon_{\mu\nu\rho\sigma} \int d\tau_r {d y_r^\sigma\over d\tau_r} 
(\tau_r) \delta^4 (x - y_r (\tau_r)),
\eeq
which are PD--dual to the support of the worldlines in ${\bf R}^4$, 
i.e. $J_{(r)}$ = PD $(\gamma_r)$.
The total electric and magnetic currents generated by the dyons are given 
by \footnote{The minus sign in the definition of $J_2$ is due to our 
conventions for differential forms. The differential $d$ acts from the
right, $d(\phi_p\wedge \phi_q)=\phi_p \wedge d\phi_q +(-)^q d\phi_p\wedge
\phi_q$, and the components of a form are defined by $\phi_p={1\over p!}
dx^{\mu_1}\wedge\cdots\wedge dx^{\mu_p}\phi_{\mu_p\cdots\mu_1}$.}
\beq\label{2.4}
J_1 = \sum_r e_r J_{(r)}, \quad J_2 = -\sum_r g_r J_{(r)}. 
\eeq
The electric and magnetic currents for each species are individually 
conserved and this is expressed through the equations
\beq\label{2.5}
dJ_{(r)} = 0. 
\eeq
Since ${\bf R}^4$ is contractible, one can apply the Poincar\`e  lemma (for
currents \cite{16}) 
and \eref{2.5} implies the existence of a 2--current $C_{(r)}$ 
such that $J_{(r)} = dC_{(r)}$. Actually one can choose $C_{(r)}$ in a
more specific form: from (2.5) and $J_{(r)} =$ PD $(\gamma_r)$ one derives
that $\gamma_r$ is the boundary of a 2--surface $\Sigma_r$ and we can 
choose 
\beq\label{2.6}
C_{(r)} = {\rm PD} (\Sigma_r).
\eeq
We then set
\beq\label{2.7}
C_1 = \sum_r e_r C_{(r)}, \quad C_2 = -\sum_r g_r C_{(r)}, 
\eeq
and we obtain
\beq\label{2.8}
J_1 = dC_1, \quad J_2 = dC_2. 
\eeq
If $\gamma_r$ describes the world line of a magnetic pole, one can 
think of $\Sigma_r$ as the support of the world--surface swept out by 
its Dirac string.

Within this set up the classical action proposed by Schwinger \cite{12} 
to derive the
Maxwell--Dirac equations \eref{1.3} is given by
\beq
\label{2.9}
S_{S} (A, J_1, C_2)= \int {1\over 2} (dA + C_2) \wedge * (dA +C_2) + 
A\wedge  J_1,
\eeq
where $A$ is a 1--form describing the electromagnetic gauge potential. 
Independence of the choice of the ``Dirac--string" $C_{(r)}$, satisfying 
$dC_{(r)} = J_{(r)}$, is a consistency condition for the theory described by
the action \eref{2.9}.

At the quantum level  the effective action $S_{eff}$ is defined by
\beq
e^{iS_{eff} (J_1, C_2)} = \int {\cal D} A e^{i S_{S} (A, J_1, C_2)}.
\eeq
A simple calculation proves that
\beq
\label{2.10}
S_{eff} (J_1, C_2) = \int -{1\over 2}\left(J_1 \wedge *  \quadratello^{-1} 
J_1+J_2 \wedge * \quadratello^{-1} J_2\right) + J_1 \wedge \delta 
\quadratello^{-1} C_2, 
\eeq
where $\quadratello = \delta d+d \delta$ is the D'Alambertian and $\delta = 
* d *$ is the codifferential.
Consistency requires that $e^{i S_{eff} (J_1, C_2^\prime)} = e^{i S_{eff} 
(J_1, C_2)}$, where $C_2^\prime$ is a new Dirac--string, i.e. the 
exponentiated effective action has to be independent of the choice of
Dirac--string. The new Dirac--strings are represented by surfaces
$\Sigma^\prime_r$, whose boundaries are again $\gamma_r$.
We can always write
\beq
C_{(r)}^\prime = C_{(r)} + d H_{(r)},
\eeq
where $H_{(r)} =$ PD $(V_r)$, $V_r$ being a 3--volume bounded by 
$\Sigma_r^\prime - 
\Sigma_r,$ and $C^\prime_{(r)} =$ PD $(\Sigma^\prime_r)$. This leads to
\bea
\nonumber
C^\prime_2&=&C_2+dH_2\\
\label{shift}
H_2&=&-\sum_r g_r H_{(r)}.
\eea
From \eref{2.10}, using $d J_1 =0$, one obtains
\beq
\label{2.11}
S_{eff} (J_1, C_2^\prime)-S_{eff} (J_1, C_2)=
\int J_1 \wedge \delta \quadratello^{-1} dH_2 = \int J_1 \wedge  H_2 =
-\sum_{r,s} e_r g_s \int J_{(r)} \wedge H_{(s)}. 
\eeq
Since $\int J_{(r)}\wedge H_{(s)}$ is integer, consistency is achieved if
\beq
\label{2.12}
e_r g_s =2\pi n_{rs}, 
\eeq
i.e. if Dirac's quantization condition holds.
This turns out to be also the condition to implement the ${\bf
Z}_4$--duality group at the quantum level. In fact the generator of ${\bf
Z}_4$ maps $S_{eff} (J_1, C_2)$ to $S_{eff} (J_1, C_2) + \int C_1 
\wedge C_2$, see \cite{1}. The last term is irrelevant if
\beq
\int C_1 \wedge C_2 = \sum_{r,s} e_r g_s \int C_{(r)} \wedge C_{(s)} 
\in 2\pi {\bf Z}.
\eeq
Since the integrals are integer this condition is fulfilled thanks to 
\eref{2.12}.

\subsection{Quantum field theory}

Next we extend these ideas to a quantum field theory setting. 
Following an intuition due essentially to Zwanziger \cite{8,13}, the key 
ingredient of the consistency check of a QFT of dyons is a 
representation 
of correlation functions of local observables in terms of Feynman path
integrals over closed classical currents, coupled to the gauge field.
This procedure corresponds to a realization of 
quantum field--particle duality.

In order to be selfcontained  and to set up the
notations, let us give an example of this procedure in a simplified model: 
a complex scalar field $\phi$ of charge $e$ coupled to the gauge field $A$, 
without additional interactions. The action of the system is given by 
$S(A)+ S(A,\phi)$, where
\beq\label{2.13}
S(A, \phi) = -\int d^4x\,  \bar\phi (\quadratello_{eA} + m^2) \phi,
\eeq
$\bar\phi$ denotes the complex conjugate field and 
$\quadratello_{eA}$ is the covariant D'Alambertian. $S(A)$ denotes the 
free Maxwell action.
Consider  the gauge--invariant correlation function
\beq
\langle T\,\phi (x)\,e^{ie\int j_{xy} \wedge A}\, \bar\phi (y) \rangle (A),
\eeq
where $\langle \, \rangle (A)$ denotes the expectation value with respect to 
the ``measure" in the path--integral induced by $S(A, \phi)$, with $A$ treated 
as an external field. $j_{xy}$ is the PD of a curve joining $x$ to $y$.

The relevant Feynman--Schwinger representation can be 
derived formally as follows:
\bea
\nonumber
\langle T \,\phi (x) \bar\phi (y) \rangle (A) &=& 
(\quadratello_{eA} + m^2 + i\varepsilon)^{-1} (x,y)\\
&=&
 i \int^\infty_0 d s\, e^{is (m^2 +i\varepsilon)} (e^{-is 
\quadratello_{eA}}) (x,y).
\label{2.14}
\eea
The operator $\quadratello_{eA}$ can be regarded as the ``Hamiltonian" of 
a particle
of mass ${1\over 2}$ and charge $e$ in 3+1 dimensions, minimally coupled to 
the gauge field $A$.
The last term in \eref{2.14} can then be viewed as the evolution kernel of this 
particle with initial position $x$ at ``time" $0$  and final position $y$ 
at ``time" $s$. It allows, therefore, a Feynman path integral 
representation:
\beq
\label{2.15}
\bigl(e^{-is\quadratello_{eA}}\bigr) (x, y) = \int {\cal D} y (\tau)\, e^{i
\int^s_0 d\tau  \bigl[{1 \over 4} {\dot y^2 (\tau)}
+ e \dot y_\mu (\tau) A^\mu (y(\tau)) \bigr]}.
\eeq
Associating to the trajectory $\{y (\tau)\}$, which starts
from $x$ and ends in $y$, a current $J$ as in \eref{2.3} one 
can write for a suitable ``measure" ${\cal D}\mu (J)$:
\beq
\label{2.16}
\left\langle T \, \phi (x) e^{ie\int j_{xy}\wedge A} \bar\phi (y)
\right\rangle (A)
= 
\int {\cal D} \mu (J)\, e^{i e\int (j_{xy} + J)\wedge A}.
\eeq
Notice, in particular, that $d(j_{xy}+J)=0$.
Similarly for the partition function $Z(A)$ of the field $\phi$ one can 
write:
\bea
\nonumber
Z(A) &=& \int {\cal D} \phi\, e^{-i\int \bar\phi (\quadratello_{eA} 
+ m^2 + i\varepsilon) \phi} 
= {\rm det}^{-1} (\quadratello_{eA} + m^2 + i\varepsilon) \\
&=&
{\rm exp}\left[-\int d^4 x \int^\infty_0 {ds\over s}\, e^{i(m^2 + 
i\varepsilon)s} \bigl(e^{-i s\quadratello_{eA}}\bigr) (x,x)\right]
= \int {\cal D}\mu ({\bf J}) e^{i e \int A\wedge {\bf J}},
\label{2.17}
\eea
for a suitable ``measure" ${\cal D}\mu ({\bf J})$ on networks ${\bf J}$ 
of {\it closed} currents. 

Finally for the normalized correlation functions we have
\beq
\label{2.18}
\left\langle T \phi (x) e^{i e \int j_{xy}\, \wedge A} \bar\phi (y)
\right\rangle
= {\int {\cal D} A\, e^{iS(A)} \int {\cal D} \mu ({\bf J})\, {\cal D} \mu 
(J)\, e^{i e \int (j_{xy} + J + {\bf J}) \wedge A} \over \int 
{\cal D} A\,
e^{iS(A)} \int {\cal D}\mu ({\bf J})\, e^{i e \int {\bf J} \wedge A}},
\eeq
which is the representation we will need for our purposes.

We present now the construction of a consistent quantum field 
theory of ${\bf Z}_4$--dyons, where the dyon fields are a family of 
complex scalars 
$\{\phi_r\}$ with charges $\{e_r, g_r\}$, interacting with a gauge
field $A$.
The basic idea is to start from the Schwinger action \eref{2.9} and to promote
$C_2$ to a real field variable, obeying the constraint 
\beq
\label{2.19}
dC_2 = J_2 (\{\phi_r\}, A), 
\eeq
where $J_2 (\{\phi_r \}, A)$ is the Hodge dual of the
total magnetic current generated by the 
fields $\{\phi_r \}$, i.e. of $-i\sum_r g_r\bar \phi_r D_r^\mu\phi_r+c.c.$
The covariant derivative appearing here will be specified below.
Eventually we will apply field/particle duality to prove the
consistency of the theory, provided Dirac's quantization condition 
\eref{2.12} holds.

The problem related with the constraint \eref{2.19} is that it does
not specify completely the field $C_2$: this constraint determines $C_2$
only modulo exact forms. To determine this field completely
we proposed in \cite{1} to modify the Schwinger action as 
follows: one introduces a constant vector $u^\mu$ satisfying $u^2\neq 0$
\footnote{Actually one can use a generic  
nowhere lightlike vector field $U^\mu (x)$, see \cite{1} for this more 
general case.} and the Lagrange multiplier fields $A_1$, a
real 1--form, and $C_1$, a real 2--form.
Setting $A\equiv A_2$ we define the QFT Schwinger action, which depends
also on the constant vector $u$, as
\bea
\nonumber
S^{u}_S (A_1, A_2, C_1, C_2, \{\phi_r \}) &=&
\int {1\over 2} (dA_2 + C_2) \wedge *(dA_2 + C_2) + A_1 \wedge d C_2
- C_1\wedge u i_u C_2\\
&& - \sum_r \int d^4x\,\bar\phi_r 
(\quadratello_{e_r A_2 + g_r A_1} + m^2_r) \phi_r, 
\label{2.20}
\eea
where $ui_u$ denotes the projection of the 2--form $C_2$ 
along $u$; in components $(ui_u C_2)
_{\rho\sigma}= 2u_{[\rho} u^\beta (C_2)_{\beta\sigma]}$. 
The covariant derivative on the $r$--th dyon field is defined by
$D^\mu_r=\partial ^\mu+i(e_r A_2^\mu + g_r A_1^\mu)$.
Furthermore we assume vanishing boundary conditions for $C_1$ and $C_2$ 
at $x^\mu u_\mu = -\infty$.

As shown in \cite{1},
the equations of motion (and symmetries of the action) determine the 
auxiliary fields in terms of $A_2$ and $\{\phi_r\}$ as follows:
\bea
\nonumber
dA_1 + C_1 &=& * (dA_2 + C_2),\nonumber\\
(\quadratello_{e_r A_2 + g_r A_1} + m^2_r)\phi_r&=&0\nonumber\\
i_u C_1 &=& i_u C_2 =0,\nonumber\\
dC_1&=& J_1  (\{\phi_r\}, A_1, A_2),\nonumber\\ 
d C_2& =& J_2  (\{\phi_r\}, A_1, A_2).
\label{2.21}
\eea
Here $J_2(\{\phi_r\}, A_1, A_2)$ is defined as the Hodge--dual of 
$i\sum_r e_r\bar \phi_r D_r^\mu\phi_r+c.c.$
The first equation is the standard duality relation $F_1=*F_2$, which 
allows to eliminate $A_1$ and implies the Maxwell equation $d*F_2=J_1$.
The second equation is the covariant Klein--Gordon equation for the
matter fields, and the remaining equations determine $C_1$ and $C_2$
completely, see below. Therefore, there are no 
unphysical propagating degrees of freedom.

It remains to prove that gauge--invariant correlation functions are 
independent of the choice of $u^\mu$. We exhibit the proof for the 
partition function; for generic correlators see \cite{1}.
As in eq. \eref{2.17} we can write:
\beq\label{2.22}
\int \prod_r {\cal D} \phi_r 
{\rm exp} \left[-i\int d^4x\,\bar\phi_r (\quadratello_{e_r A_2 + g_r A_1}
+ m^2_r)\phi_r\right]
= \int \prod_r {\cal D} \mu ({\bf J}_{(r)}) e^{i\int J_1 \wedge A_2 - 
J_2 \wedge A_1},
\eeq
where
\beq\label{2.23}
J_1 = \sum_r e_r {\bf J}_{(r)}, \quad J_2= -\sum_r g_r {\bf J}_{(r)}, 
\eeq
and the {\bf J}$_{(r)}$ represent a network of closed currents 
corresponding to the dyon field $\phi_r$. 

Using \eref{2.22} and integrating out $A_1$ and $C_1$ one can write the 
partition function $Z_u$, associated to the action \eref{2.20}, as
\bea\nonumber
Z_u &=& \int {\cal D} A_2 {\cal D} C_2 \prod_r {\cal D} \mu ({\bf J}_{(r)}) 
\,e^{i\int{1\over 2} (dA_2 + C_2) \wedge * (dA_2 + C_2)}\\
&&
\delta (i_u C_2)\, \delta (dC_2 - J_2)\, e^{i\int J_1\wedge A_2}. 
\label{2.24}
\eea
Together with the boundary condition along 
$x^\mu u_\mu \rightarrow -\infty$, the two constraints appearing in 
\eref{2.24} fix $C_2$ uniquely as
\beq\label{2.25}
C_2 (u, J_2) = (u^u \partial_\mu)^{-1} i_u  J_2. 
\eeq
In order to satisfy the boundary condition the kernel $G$ associated to  
the inverse operator $(u^\mu 
\partial_\mu)^{-1}$ has to be defined as 
\beq\label{2.26}
G(x) = \Theta (u_\mu x^\mu) \delta^3 (\vec x_u^\bot), \qquad  u^\mu
\partial_\mu G (x) = 
\delta^4 (x), 
\eeq
where $\vec x^\bot_u$ are the three coordinates orthogonal to 
$u_\mu x^\mu$. The explicit expression of $C_2$ is given by
\beq\label{2.27}
C_2(u, J_2)=-\sum_r g_r C_{(r)},
\eeq
where
\beq
\label{2.28}
C_{(r)} (x) = {1\over 2}\, dx^\mu \wedge  dx^\nu \varepsilon_{\mu\nu\rho\sigma} 
u^\rho
\int^\infty_0 ds \int^{+\infty}_{-\infty} d\tau_r {dy^\sigma_{(r)}\over 
d\tau_r} (\tau_r) \delta^{(4)} (x - (y_{(r)} (\tau_r) + u s)),
\eeq
and $y_{(r)}$ parametrizes the network of worldlines supported on the 
set of curves ${\bf \gamma}_{(r)}$,
corresponding to the current network ${\bf J}_{(r)}$.
The geometric interpretation of $C_{(r)}$ is rather simple.
It is PD of a surface $\Sigma_{(r)}(u)$ whose 
boundary is ${\bf \gamma}_{(r)}$ and whose generators are all parallel to 
$u^\mu$. These two requirements specify $\Sigma_{(r)}(u)$ completely. 
$C_{(r)}$ is, in particular, an integer form.

The integration over $C_2$ in \eref{2.24} becomes therefore trivial, and
the final integration over $A_2$ has already been performed for the
original Schwinger action and led to the result \eref{2.10}. 
Putting everything together we obtain
\beq
\label{2.29}
Z_u = \int \prod_r {\cal D}\mu({\bf J}_{(r)} ) e^{iS_{eff} (J_1, C_2 
(u, J_2))}. 
\eeq
For a different vector $u^\prime$, since
\beq
dC_2 (u, J_2) = dC_2 (u^\prime, J_2)= J_2,
\eeq
one can write
\beq
\label{2.30}
C_2 (u^\prime, J_2) = C_2 (u, J_2) + dH_2(u, u^\prime),
\eeq
where the 1--form $H_2(u, u^\prime)$ is a linear combination
of  the Poincar\`e  Duals of 
3--volumes bounded by the surfaces $\Sigma_{(r)} (u) - \Sigma_{(r)} 
(u^\prime)$, weighted by the magnetic charges $g_r$, see \eref{shift}.
It is then clear that by the same mechanism acting in eq. \eref{2.11}, the 
partition function is independent of the choice of $u$, provided Dirac's 
quantization condition \eref{2.12} holds.

The same strategy applies to all correlation functions of local (neutral) 
gauge--invariant observables. 

\subsection{Reflection positivity}

The analysis developed until now was based 
on a Minkowskian formalism, but it is easy to perform a transition to a
euclidean formalism. In particular, the Schwinger action $S_{S}^u$ has to 
be replaced 
with its euclidean counterpart, obtained multiplying by $i$ the second and 
third terms in \eref{2.20}, and using everywhere the euclidean metric.

Starting from the full set of euclidean correlation 
functions of local gauge--invariant observables, provided a set of 
properties (the Osterwalder--Schrader (O.S.) axioms \cite{17}) 
are satisfied, one can 
reconstruct the Hilbert space of states of the vacuum sector, ${\cal H}_0$,
containing the vacuum state $|\Omega>$, carrying a unitary
representation of the 
(covering of the) Poincar\`e group, $\tilde{\cal P}^\uparrow_+$ 
leaving $|\Omega>$ invariant, and quantum field operators corresponding to the 
(classical) euclidean fields.   
Therefore, the full
structure of a Relativistic Quantum Field Theory in its vacuum sector can 
be reconstructed out of the euclidean correlation functions of local 
observables, provided O.S. axioms hold.

There is also a version of this reconstruction theorem that applies to 
lattice regularized theories \cite{18}.

The two basic O.S. axioms which allow to set up the entire 
formalism mentioned above are: 

-- invariance of the euclidean correlation functions under the euclidean 
group (lattice translations, in the lattice) 

-- reflection (or O.S.) positivity. In the models considered
here reflection 
positivity can be  defined  as 
follows: let ${\cal F}_+$ denote 
the algebra of gauge invariant functions of the euclidean fields, i.e. 
euclidean observables, with
support in the positive time 4--space; let 
$\Theta$ denote reflection w.r.t. the time zero space followed by complex
conjugation. Then reflection positivity means that $\forall{\cal F}\in 
{\cal F}_+$

\beq
\langle {\cal F} \Theta {\cal F} \rangle \geq 0,
\eeq
where $\langle \, \rangle$ denotes the euclidean expectation value.

The relation between euclidean observables and quantum observables is then
given as follows: let ${\cal A}_+ \subset {\cal F}_+$ denote the 
polynomial algebra generated by euclidean local observables supported
in the positive time 4--space. To each element ${\cal O} \in {\cal A}_+$ 
we associate a vector $\vert{\cal O}\rangle \in {\cal H}_0$.
The scalar product between such vectors is given by
\beq
\langle{\cal O} \vert {\cal O}^\prime \rangle = 
\langle {\cal O}^\prime \Theta
{\cal O} \rangle.
\eeq
Let ${\cal O}= \prod_i {\cal O}_{x^0_i},$ where ${\cal O}_{x^0_i}$ is
an euclidean local observable with support at fixed time $x^0_i$.
Then, for $0\leq x^0_j < x^0_{j+1}$, we have 
\beq
\vert {\cal O} \rangle = \prod_i \hat{\cal O}_{x^0_i} \vert
\Omega \rangle, 
\eeq
where $\hat{\cal O}_{x^0_i}$ is the quantum field operator corresponding
to the (classical) euclidean field ${\cal O}_{x^0_i}$. If we denote by $H$ the 
Hamiltonian, generator of time translations, then formally 
$\hat{\cal O}_{x^0} 
= e^{- x^0 H} \hat{\cal O}_0 e^{x^0 H}$, where $\hat{\cal O}_0$ is a 
standard "time--zero quantum field operator".
[An analogous
relation holds between euclidean charged fields and quantum charged
fields discussed in the next section: for a more detailed discussion
of these methods see \cite{11} and references therein].

Let us show that at a formal level the QFT of dyons defined above 
satisfies euclidean invariance and reflection positivity.
For the euclidean ${\bf Z}_4$--theory of dyons  
invariance under the euclidean group
has been established above; in fact, the action $S_S^u$ is  manifestly 
invariant 
under the euclidean group and the dependence on the fixed 
vector $u^\mu$ has been shown to be spurious at the quantum level.
  
Reflection positivity can be proved by standard arguments \cite{18}.
We use gauge invariance to set the temporal component of the gauge fields
to zero. Furthermore we choose $u_\mu$ along the time direction, 
$u_\mu=(1,\vec 0)$. Integration over $C_1$ sets then $C_2^{0i}=0$
and the integration over $C_2$ reduces to an integration over 
$C_2^{ij}\equiv c^{ij}$. The integration measure becomes then
${\cal D} {\cal M} \equiv {\cal D} \vec A_1 {\cal D} \vec A_2 
{\cal D}c \prod_r {\cal D} \phi_r$. Setting 
$$
\widetilde S(\vec A_1,\vec A_2,c,\{\phi_r \})
\equiv S^{u}_S (A_1, A_2, 0, C_2, \{\phi_r \})\Bigg|_{A_1^0=A_2^0=
C_2^{0i}=0},
$$
schematically we have:
\bea
\langle {\cal F}
(A_1, A_2, C_2)\Theta {\cal F} (A_1, A_2, C_2) \rangle&=&
\int {\cal D}{\cal M} 
\, e^{-\widetilde S}\, 
{\cal F} (\vec A_1, \vec A_2, c) \,\Theta {\cal F} 
(\vec A_1, \vec A_2, c)\nonumber\\
&=&\int {\cal D}{\cal M}\, 
e^{-\widetilde S_{x^0>0}}\, {\cal F} (\vec A_1,\vec A_2,c)\,
\Theta \left[e^{-\widetilde S_{x^0>0}}\,  
{\cal F} (\vec A_1, \vec A_2, c)\right]\nonumber\\
&=&
\left|\int_{x^0 > 0}
{\cal D} {\cal M}\, 
e^{- \widetilde S_{x^0> 0}} {\cal F} (\vec A_1, \vec A_2, c)
\right|^2 \geq 0.
\nonumber
\eea
This computation can be made mathematically precise with a lattice 
regularization. 

\subsection{SO(2)--Dyons}

Let us briefly turn to the quantum field theory of $SO(2)$--dyons. 
With the same notation 
adopted for the {\bf Z}$_4$--theory, the (manifestly) $SO(2)$--invariant  
action proposed by  Zwanziger \cite{8} for
the Dirac--Maxwell equations \eref{1.3}, is constructed as follows.

First of all, to realize the  $SO(2)$--duality group as a manifest
invariance, one introduces a doublet of vector potentials 
$A\equiv {A_1\choose A_2}$. Introducing also 
a constant 4--vector $n^\mu$ with $n^\mu n_\mu=-1$, one constructs a 
$2\times 2$ matrix--valued operator $Q(n)$ which sends 2--forms into 2--forms
\beq
\label{2.32}
Q(n) = \left(\matrix {*n i_n & {1\over 2} -  n i_n\cr
-{1\over 2} + n i_n & *n i_n\cr}\right). 
\eeq
The action is then given by
\beq
\label{2.33}
S^n_Z (A_1, A_2, J_1, J_2)= \int {1\over 2} (dA)^T \wedge Q (n) dA + A_1 
\wedge J_2 - A_2 \wedge J_1,
\eeq
where $(\cdot)^T$ denotes transposition, and one assumes
the boundary conditions
\beq
A_\alpha (n^\mu x_\mu = -\infty) = 0,\alpha= 1,2.
\eeq
The effective action turns out to be 
\beq
\label{2.34}
S_{eff}^n = - {1\over 2} \int  J_\alpha \wedge*\quadratello^{-1} J_\alpha -
\varepsilon^{\alpha\beta} J_\alpha \wedge  \quadratello^{-1}\delta C^n_\beta,
\eeq
where summation over the $SO(2)$ indices $\alpha,\beta$ is understood, and 
the 2--forms $C^n$ (Dirac--strings) are given by
\beq
\label{2.35}
C^n_\beta = (n^\mu \partial_\mu)^{-1} i_n J_\beta,
\eeq
where the inverse operator $(n^\mu \partial_\mu)^{-1}$ is defined as in 
equation \eref{2.26}. 

The difference between the $SO(2)$-- and ${\bf Z}_4$--theories is clearly
exhibited by the corresponding effective actions, \eref{2.34} and 
\eref{2.10}. 
Choosing $n^\mu=u^\mu$ one has indeed
$$
\left(S_{eff}^n\right)_{{\bf Z}_4}-\left(S_{eff}^n\right)_{SO(2)}
=  {1\over2} \int\left(J_1 \wedge \delta\quadratello^{-1} C_2
+J_2 \wedge \delta\quadratello^{-1} C_1\right)=
{1\over2}\int C_1 \wedge C_2,
$$
where the last step, following from Hodge decomposition of the
Laplacian, is formal because of self--interactions which need
a regularization, see below. It is eventually this difference which
leads to spin--statistics transmutation in the $Z_4$--theory, but not in
the $SO(2)$--theory.

$C^n$ describes again the (time--evolution of the) Dirac--string,
directed along $n$, and the consistency requirement is therefore 
the independence of the
exponentiated effective action of the choice of $n$ \footnote{In
Zwanziger's classical action the vector $n$ induces the projection
operator $Q(n)$ in the kinetic terms for the gauge fields and has nothing 
to do with the Dirac--string. Only at the quantum level, i.e. in 
$S_{eff}^n$, it acquires the meaning of the direction of the Dirac--string
and, only then, it can be identified with $u$.}. 
Since $dC^n_\alpha =
J_\alpha$, for a different unit vector $n^\prime$ we have
\beq
C^{n^\prime}_\alpha = C^n_\alpha + d H_\alpha(n,n^\prime),
\eeq
where the 1--forms $H_\alpha(n,n^\prime)$ are linear combinations of
Poincar\`e Duals of 3--volumes as in eq. \eref{2.30}.
A simple computation gives
\bea
\nonumber
S^{n^\prime}_{eff}-S^n_{eff}=
{1\over 2} \int \varepsilon^{\alpha\beta} J_\alpha \wedge \delta  
\quadratello^{-1} dH_\beta(n,n^\prime)&=& {1\over 2} 
\int \varepsilon^{\alpha\beta}
J_\alpha\wedge H_\beta(n,n^\prime)\\
&=& {1\over 2}  \sum_{r,s} (e_s g_r - g_s e_r) \int J_{(r)} \wedge 
H_{(s)}(n,n^\prime),\nonumber
\label{2.36}
\eea
and since $\int J_{(r)} \wedge H_{(s)}(n,n^\prime) \in$ {\bf Z}, one 
obtains as consistency condition the Schwinger--Zwanziger quantization 
condition
\beq
\label{2.37}
{1\over 2} (e_r g_s - g_r e_s) \in 2\pi {\bf Z}.
\eeq

From the explicit form of $S^n_{eff}$, eq. \eref{2.34}, it is clear that 
also at the quantum level the $SO(2)$--duality group is realized as a manifest
symmetry, once independence of $n$ has been established.

Since Zwanziger's classical action, eq. \eref{2.33}, involves only the 
currents (and not the strings $C$), the 
transition to  quantum field theory  is straightforward. 
Zwanziger's action for quantum dyon fields $\{\phi_r\}$ reads:
\beq
\label{2.38}
S^n_Z (A_1, A_2, \{\phi_r\})= \int {1\over 2} (dA)^T \wedge Q (n) dA
- \sum_r \int d^4x \,\bar \phi_r (\quadratello_{e_r A_1 + g_r A_2} + m^2_r) 
\phi_r.
\eeq
The proof of independence of $n$ of correlation functions of local 
gauge--invariant observables can be achieved using representations in 
terms of closed currents, as in {\bf Z}$_4$--theories, provided the 
Schwinger--Zwanziger quantization condition holds.

Again, one can obtain a euclidean formulation for $SO(2)$--theories by 
replacing $S^n_Z$ with a euclidean action, obtained multiplying 
the off--diagonal terms in $Q (n)$ in \eref{2.38} by $i$ and 
using the euclidean metric.                           

At a formal level euclidean invariance of local observables is ensured, 
once their independence of the choice of $n$ has been established; 
O.S. positivity can be proven with tools similar to the ones used
in {\bf Z}$_4$--theories, choosing the vector $n^\mu$ with vanishing 
time component and setting  $A^0_\alpha =0$, using gauge invariance. 

\section{Gauge--invariant charged fields}

The proof of unobservability of the Dirac--string 
in  correlation functions of local 
observables depends crucially on their representation in terms of 
{\it closed} Feynman paths. Closed paths amount, via PD, to 
conserved currents and current conservation is, in turn, a consequence of
invariance under gauge transformations.  Gauge invariance is 
therefore a natural request in the construction of correlation functions 
for charged fields. This requirement can be fulfilled by means of an ansatz
due to Dirac \cite{10}. To illustrate the ansatz (in its euclidean 
formulation)
and to explain why it has to be modified in a QFT of dyons, we  
exemplify it in the simple model of a complex scalar field, coupled to a 
gauge field, discussed previously in eqs. \eref{2.13}--\eref{2.18}.

Let $E=dy^\mu E_\mu(y)$ denote a 1--form, with support in the time zero 
3--space, satisfying $\partial_\mu E^\mu=\delta^4 (y)$. More precisely,
\beq
\label{3.1}
E^0(y) =0, \quad E^i(y) = \delta(y^0)E^i (\vec y),\quad
\partial_i E^i (\vec y) = \delta^3 (\vec y),\quad i=1,2,3
\eeq
with
\beq
\label{3.2}
|\vec E (\vec y)| = O\left({1\over |\vec y|^2}\right),  
\quad {\rm for}\,\,|\vec y| \rightarrow \infty. 
\eeq
In particular, a rotation--symmetric choice of $\vec E$ corresponds to the
(classical) electric field generated by a unitary charge located at 
the origin, $\vec E(\vec y)=\vec y/4\pi |\vec y|^3$.
One may, however, consider different choices of $\vec E$ corresponding to 
anisotropic spreadings of the electric flux generated by the charge, which 
fulfill still the decreasing condition \eref{3.2}. In particular, one can 
concentrate all the flux inside a cone ${\cal C}$ with apex at the origin. This 
choice will be relevant in later discussions.

We denote by $E^x$ the 3--form hodge--dual to the above defined 1--form 
$E$, translated by $x$, and define the charged fields 
\bea
\nonumber
\phi (E^x) &=& \phi (x) e^{i e\int E^x \wedge A}\\
\bar\phi (E^y) &=&\bar\phi (y) e^{-i e \int E^y \wedge A}.
\label{3.3}
\eea
Euclidean correlation functions of charged fields are then given by
$$
\left\langle \prod^n_{i=1} \phi (E^{x_i}) \prod^n_{j=1} \bar\phi (E^{y_j}) 
\right\rangle,
$$
and by charge conservation they vanish if the numbers of fields $\phi$ and 
$\bar\phi$ are different.

At a formal level one can show that the (mixed) euclidean correlation 
functions of charged fields and neutral observables defined according to the 
above prescription satisfy a variant of O.S. axioms, in particular  
translation invariance and 
reflection positivity. These  axioms allow then to reconstruct a Hilbert 
space of physical states labelled by $E$, ${\cal H} (E)$,  
carrying a unitary representation of translations, $U_E$. They allow
also to reconstruct quantum 
charged field operators $\hat\phi (E^x), \hat{\bar\phi}(E^y)$ (with 
charges $+e$ and $-e$ respectively) and quantum observables acting on 
${\cal H} (E)$.

Furthermore, from the vanishing of correlation functions of non--zero total 
charge, it follows that the Hilbert space ${\cal H} (E)$ splits into a 
direct sum of subspaces ${\cal H}_q (E)$ with fixed electric charge $qe$,
($q\in{\bf Z}$):
$$
{\cal H} (E) = \oplus_q {\cal H}_q (E).
$$

All these formal considerations  can be made 
mathematically precise in a lattice--regularized version \cite{19}.

One may ask if the Hilbert spaces ${\cal H}_q (E)$ and ${\cal H}_q 
(E^\prime)$ corresponding to different choices of the 1--form $E$ are 
orthogonal to each other. 
To analyse this problem one should consider correlation functions of the 
fields $\phi (E)$, $\phi (E^\prime)$ and their complex conjugates. 
Consider e.g. the two point function 
$\langle \phi (E^x) \bar\phi(E^{\prime
y}) \rangle$. In a lattice regularization one can show \cite{19}  
that, as $|x-y| \rightarrow \infty$ it behaves like
\beq
\label{3.4}
\langle \phi (E^x) \bar\phi(E^{\prime
y}) \rangle\sim
e^{-c \int (E^x-E^{\prime y}) \wedge * \Delta^{-1} (E^x -E^{\prime y})} 
\cdot{e^{-c^\prime |x-y|} \over |x-y|^{3/2}},
\eeq
where $c$ and $c^\prime$ are suitable positive constants and
$\Delta$ is the 4--dimensional laplacian. Hence, the correlation 
function vanishes at large distances for electric fields $E$ and
$E^\prime$ for which the integral
$$
\int (E^x - E^{\prime y}) \wedge * \Delta^{-1} (E^x - E^{\prime y})
$$
diverges. This happens if the behaviour at infinity of $\vec E$ is different 
from that of $\vec E^\prime$, e.g. if $\vec E$ is supported in a cone 
${\cal C}$ and $\vec E^\prime$ is rotation symmetric or supported in a cone 
${\cal C}^\prime$ not overlapping with ${\cal C}$.
A little elaboration shows that if $\int (E-E^\prime)* \wedge \Delta^{-1} 
(E-E^\prime)$ diverges, then ${\cal H}_q (E) \bot {\cal H}_q (E^\prime)$. 

This construction of charged fields, based on Dirac's ansatz, breaks down if
electric and magnetic dynamical charges coexist, as in QFT of dyons, due to
the consistency requirement of unobservability of the Dirac--string.
This failure can be immediately understood by noticing that in the effective
action \eref{2.10}, corresponding to a QED with electric {\it and} 
magnetic currents, in euclidean space--time the last term reads 
\beq
\label{3.5}
i\int J_1 \wedge \, \delta \Delta^{-1} C_2.
\eeq
Using a representation in terms of currents analogous to \eref{2.18}, 
one can see 
that the electric current $J_1$, appearing in the correlation functions 
of charged
fields constructed according to Dirac's ansatz, involves also  
contributions due to the ``classical electric fields" $E$.
For example, in the correlator $\langle\phi (E^x) \bar\phi (E^y) 
\rangle$ the electric current 3--form, appearing in the effective action,
is given by
$J_1= J+ e(E^x - E^y)$, where $J$ is associated, via PD, to an open curve
with boundary $\{x, y\}$; therefore the total current is again 
conserved, $d(J + e (E^x - E^y)) =0$.
But the effective action exhibits now a term $\int [J + e(E^x - E^y) 
] \wedge \, \delta \Delta^{-1} C_2$, which is obviously not invariant under a
change of Dirac--string, $C_2\rightarrow C_2+dH_2$, because:
\beq
\label{pippo}
\int (e(E^x - E^y) + J) \wedge \delta \Delta^{-1} dH_2 = \int [e(E^x - 
E^y) + J] \wedge H_2 \not\in 2\pi {\bf Z}.
\eeq
In other words, this formulation is inconsistent because the 
3--current $E$ is {\it not an integer current}; its integrals over arbitrary
manifolds, contrary to what happens for point--like currents $J$, do not 
belong to ${\bf Z}$. Therefore, Dirac's quantization condition is not
sufficient to make the variation \eref{pippo} an integer multiple of $2\pi$.

A naive arrangement to avoid this inconsistency would be the replacement 
of $E^x$ by a ``Mandelstam string"
$\gamma^x$, a 3--form current which is PD to a (point--like) curve starting  
from $x$ and reaching infinity, with support at fixed time $x^0$.
Such a current satisfies still
$$
d\gamma^x = \delta_x\equiv d^4y\, \delta^4(y-x),
$$
as does $E^x$, but, being an integer current, \eref{pippo} would become
an integer multiple of $2\pi$  and the Dirac--string would be unobservable.
However, what goes wrong with this recipe is that $\gamma^x$ violates
the condition \eref{3.2}, because it is a 
$\delta$--function on an infinite line. As a consequence incurable
infrared divergences would appear. 
Consider e.g. again the two--point function $\langle\phi (\gamma^x) \bar\phi 
(\gamma^y) \rangle$. From estimates like \eref{3.4} it can be shown
that every Mandelstam string $\gamma^x$ carries 
an infinite positive self--energy $(\sim \int \gamma^x \wedge * 
\Delta^{-1} \gamma^x)$, and that the interaction energy between two 
strings with opposite charges $(\sim \int \gamma^x
\wedge^* \Delta^{-1} \gamma^y)$ is infinite and negative. 
The reason for these divergences is that the Mandelstam strings have 
infinite length and that  
the corresponding electric currents do not decay sufficiently fast at 
infinity, i.e. they violate \eref{3.2}.

The diverging self--energies could be eliminated via multiplicative 
renormalization, but, since the diverging interaction terms depend on the 
distance $|x-y|$, their renormalization would spoil 
O.S. positivity, preventing the reconstruction of charged quantum fields.

A solution to these problems has been proposed in \cite{11}: 
accordingly one has to replace a
fixed Mandelstam--string $\gamma^x$ by a sum over fluctuating 
Mandelstam--strings, each one at fixed time $x^0$,
weighted by an appropriate measure 
${\cal D}\nu (\gamma^x)$.
This measure has been constructed in \cite{11} and it is 
supported on strings which fluctuate so strongly
that, with probability 1, the interaction energy between two strings 
is finite, even for an infinite length.

The two--point function for charged fields should then be defined by
\beq
\label{3.6}
\left\langle\int {\cal D} \nu (\gamma^x)\, e^{i e \int \gamma^x \wedge A} \phi 
(x) \int{\cal D} \nu (\gamma^y)\, e^{-i e {\int \gamma^y \wedge A}} 
\bar\phi (y) \right\rangle.
\eeq

It has been shown (in euclidean space--time with a lattice cutoff)
that there exists a complex measure $D\nu_E (\gamma^x)$ such that

i) the correlation functions for charged fields constructed in this
way satisfy formally the (lattice version of the) O.S. axioms

ii) at large distances, up to a multiplicative renormalization,

\beq\label{3.7}
\int {\cal D} \nu_E (\gamma^x)\, e^{i e \int \gamma^x \wedge A} \sim e^{i e 
\int E^x \wedge A},
\eeq
where $E^x$ is the ``classical" rotation invariant Coulomb field.
Hence, on large scales the fluctuating 
Mandelstam strings produce a phase factor which exhibits the same infrared 
behaviour as the one appearing in the Dirac ansatz (this has been verified 
in \cite{11} in gaussian approximation ).

By inspection of the explicit construction, one can infer that a 
straightforward modification of the recipe gives rise to measures 
${\cal D} \nu_E (\gamma^x)$ 
which at large distances behave as in \eref{3.7}, but where $E^x$ is 
an electric 
field supported in a cone ${\cal C}$ with corner in $\vec x$ 
\footnote{In the definition (4.2) of \cite{11} one has to choose Neumann 
b.c. at the boundary of ${\cal C}$.}.

From correlation functions like \eref{3.6} one can reconstruct a quantum field 
operator $\hat\phi (E^x)$ corresponding to the euclidean field 
\beq\label{3.8}
\int {\cal D} \nu_E (\gamma^x) e^{ie\int\gamma^x \wedge A} \phi (x). 
\eeq

It is clear how to adapt this construction to dyon quantum fields: we fix 
an electric field configuration $E$ satisfying \eref{3.1} and 
\eref{3.2}; to this 
configuration we associate a measure on Mandelstam strings ${\cal D} 
\nu_E (\gamma^x)$ as above. The euclidean correlation functions of the 
field
\beq
\label{3.9}
\int {\cal D} \nu_E (\gamma^x) \phi_r (x)\, 
e^{i\int {\gamma^x} \wedge \ve^{\alpha\beta} 
e_{r\alpha} A_\beta} \equiv \phi_r(E^x)
\eeq
allow (formally) the reconstruction of a quantum field operator $\hat\phi_r 
(E^x)$, acting on a Hilbert space ${\cal H}^r (E)$. It 
creates dyon states with a 
dressing cloud of soft ``photons", whose infrared behaviour is encoded by 
$E$. Here we have set $e_{r\alpha}=(e_r,-g_r)$, so that  
$\ve^{\alpha\beta} e_{r\alpha} A_\beta=e_rA_2+g_rA_1$. 

This construction holds in the  ${\bf Z}_4$-- as well as in the 
$SO(2)$--quantum field theories of dyons.

Vanishing of all correlation functions of non--zero total charge implies 
that ${\cal H}^r(E)$ splits into a direct sum of superselection sectors 
${\cal H}^r_q (E)$, $q \in {\bf Z}$, with electric charge $qe_r$ and 
magnetic charge  $qg_r$. The field operator $\hat\phi_r (E^x)$ maps the 
vacuum sector, ${\cal H}^r_0 (E)$, to ${\cal H}_1^r (E)$.

More generally, if we consider, in the models discussed here, 
correlation functions of several species 
$r=1,...,N$ of dyons, the total Hilbert space for fixed $E$ is a direct
sum of the Hilbert spaces ${\cal H}^r(E)$, because the currents associated 
to each species are individually conserved.

We can now give a functional integral representation for the two--point
function of the charged dyon fields defined in \eref{3.9}. 
For the $Z_4$--theory we use the euclidean version of the Schwinger action, 
$S_S^u$ \eref{2.20}, in the functional integral measure. Apart from 
an overall normalization we have
$$
\left\langle\bar\phi_r (E^y) \phi_r (E^x)\right\rangle=
$$
\beq
\int {\cal D} \nu_E (\gamma^y){\cal D} \nu_E (\gamma^x)
\int \prod_s {\cal D} \phi_s \prod_\alpha {\cal D} A_\alpha {\cal D} C_\alpha
\,e^{-S_S^u}\,e^{i\int(\gamma^x-\gamma^y)
\wedge \ve^{\alpha\beta} e_{r\alpha}A_\beta} \bar\phi_r(y)\phi_r(x).
\label{a}
\eeq

We saw already that the independence of the Dirac--string, i.e. of $u$,
of correlation functions is most easily 
proved in a path--integral representation; this technique applies also 
to the two--point function at hand, and we do not repeat here the relevant 
steps, since they add nothing new. However, as we will see in the next 
two sections, an analysis of spin and statistics is also most easily carried
out in such a representation. Since the spin analysis will be performed on 
the above two--point function, we give here its path--integral 
representation explicitly.

To obtain it we follow the steps outlined in section two.
First one integrates over the complex scalars according to \eref{2.18};
the role of $j_{xy}$ is here played by the 3--current $\gamma^x - \gamma^y$.
Then integration over $C_1$ and $A_1$ gives the $\delta$--functions
for $C_2$, as in \eref{2.24}. The integration over $C_2$ can then
be performed as after \eref{2.24} and fixes it in terms of the currents.
The final integration over $A_2$ gives rise to the (euclidean) effective 
action of the ${\bf Z}_4$--theory, \eref{2.10}. The result is 
\beq
\left\langle\bar\phi_r (E^y) \phi_r (E^x)\right\rangle=
\int {\cal D} \nu_E (\gamma^y){\cal D} \nu_E (\gamma^x)
\int \,{\cal D} \mu (J_{(r)}) \prod_s {\cal D} \mu ({\bf J}_{(s)})\, 
e^{-S_{eff}(j)}.
\label{3.10}
\eeq
The current 3--form doublet $j_\alpha$, $dj_\alpha=0$, is here given by
\beq
\label{jtot}
j_\alpha= \sum_s e_{s\alpha}K_s,
\eeq
with 
\bea
K_s &=&{\bf J}_{(s)}, \quad\quad {\rm for}\,\,s\neq r\no\\ 
K_r &=&{\bf J}_{(r)} +\gamma_r \label{kr}, 
\eea
where
\beq
\gamma_r\equiv\gamma^x-\gamma^y+J_{(r)}.\label{gr}
\eeq
We remember that  the ${\bf J}_{(s)}$ are closed forms of compact support, 
corresponding to closed paths in the path--integral
measure, which come from the $N$ matter determinants,
and that the current $J_{(r)}$ is associated to open paths, with endpoints
$\{x,y\}$, which comes from the insertion of the fields $\bar\phi_r(y)$
and $\phi_r(x)$. This implies that also $\gamma_r$ is a closed form, 
corresponding to a boundaryless (non compact) path which reaches infinity
along $\gamma^x$ and $\gamma^y$. 
In conclusion, the insertion of the
two--point function for the $r$--th field affects only the $r$--th 
current  $K_r$, adding a non--compact closed current.

The Dirac--string 2--forms, $C_\alpha \equiv \sum_s e_{s\alpha}C_s$,
are again completely fixed by
\bea\no
dC_s &=&K_s\\
i_u C_s &=&0\label{fix},
\eea 
for all $s$.
Explicitly, the euclidean version of the effective action \eref{2.10} 
is 
\beq
S_{eff}(j)=
\int {1\over 2}\, j_\alpha \wedge *  \Delta^{-1} 
j_\alpha +i\, j_1 \wedge \delta 
\Delta^{-1} C_2, \qquad  [{\bf Z}_4-{\rm theory}].
\label{Z}
\eeq
Notice, in particular, the appearance of the factor $i$ in the last
term which will become crucial below. By the way, from \eref{3.10} one 
sees immediately that the correlator is Dirac--string independent, because,
under a change of Dirac--string, the effective action gets shifted by an 
integer multiple of $2\pi i$, as shown previously. For this reason
we indicated as arguments of the effective action only the currents 
$j_\alpha$, and not the strings $C_\alpha$; strictly speaking it is 
the exponential $exp(-S_{eff}(j))$ which is a functional of only the 
currents.

For the $SO(2)$--theory the path--integral representation for charged
field correlators can be obtained in the same way as for the 
${\bf Z}_4$--theory. One has to use Zwanziger's 
classical action, $S^n_Z$ in \eref{2.38}, instead of Schwinger's
action in the functional integral \eref{a}, and the functional integral
measure is only over the fields $\phi_r$ and $A^\alpha$. Proceeding with 
the same steps as above one arrives to an expression which is identical
to \eref{3.10}, apart from the fact that the (euclidean) effective action 
is now the one of the $SO(2)$--theory, see \eref{2.34}:
\beq
S_{eff}(j)=
\int {1\over 2}\, j_\alpha \wedge *  \Delta^{-1} 
j_\alpha +{i\over 2}\,\ve^{\alpha\beta}j_\alpha \wedge \delta 
\Delta^{-1} C_\beta,  \qquad  [SO(2)-{\rm theory}].
\label{S}
\eeq
 
Comparing \eref{Z} with \eref{S} we remarked already that the diagonal (real) 
contributions are identical and that the difference lies entirely in the 
imaginary parts: since spin--statistics transmutation is related with
phase factors only the imaginary parts can, a priori, give rise to such a 
phenomenon. We remark also that
in the absence of magnetic charges we have $j_2=0=C_2$,
the imaginary parts disappear in both effective actions and the 
correlators reduce to the ones of ordinary scalar electrodynamics, with
$S_{eff}(j)={1\over 2}\int  j_1 \wedge *  \Delta^{-1}j_1$. Since in this
case there is no spin--statistics transmutation it is clear that also
for dyons the diagonal parts do not induce such a transmutation. 
That eventually transmutation occurs only in the ${\bf Z}_4$--theory, and
not in the $SO(2)$--theory, is related to the different structure of
the imaginary parts of the corresponding effective actions.

Notice also that in generic correlation functions each field 
$\phi_r (E)$ must be accompanied by a field 
$\bar\phi_r (\tilde E)$ corresponding to an electric distribution
$\tilde E$ with the same behaviour at infinity as $E$, because otherwise 
the correlation functions vanish, due to infrared divergences, as discussed 
previously.

\subsection{An analysis of the effective action}

We devote this subsection to an analysis of the effective actions
obtained above, since they will play a crucial role in the derivation
of spin and statistics. 

Using the above parametrizations of currents and strings we can write
the (common) real part of the effective actions as
$$
{\rm Re}\, S_{eff}={1\over 2}\sum_{s,t} (e_se_t+g_sg_t)
\int K_s \wedge *  \Delta^{-1} K_t, 
$$
while their imaginary parts can be written as
\bea\no
{\rm Im}\, S_{eff}&=&-\sum_{s,t}e_sg_t\, \Gamma(K_s,K_t)
\qquad  [{\bf Z}_4-{\rm theory}]\\
{\rm Im}\, S_{eff}&=& -{1\over2} \sum_{s,t}\,(e_sg_t-e_tg_s)\, \Gamma(K_s,K_t)
\qquad  [SO(2)-{\rm theory}].
\label{imm}
\eea
We introduced here the real bilinear functional of currents 
\beq
\label{def}
\Gamma(K_s,K_t)\equiv  \int   K_s \wedge \delta 
\Delta^{-1} C_t.
\eeq
Actually, $\Gamma$ is a functional of the currents only if it is defined 
{\it mod} {\bf Z}, because, as we saw previously, under a change  
of the string $C_t$ it changes by an integer. As $\Gamma$ changes by an 
integer, the two effective actions, each under its appropriate quantization 
condition for the charges, change by an irrelevant integer multiple of
$2\pi i$. Hence it is sufficient that $\Gamma$ is a functional of the 
currents, {\it mod} integers.

An explicit representation for it, needed below, can be obtained as 
follows. We parametrize the closed curve $l_s$ ($l_t$), associated to 
$K_s$ ($K_t$), by $x^\mu(\sigma)$ ($y^\mu(\tau)$). We choose 
$u^\mu=(1,0,0,0)$ as the direction
of the Dirac--string; for the two--form $C_t$ we can then use
the explicit expression \eref{2.28}. Since in four--dimensional euclidean 
space--time the inverse of the Laplacian, $\Delta^{-1}$, is represented by 
the Kernel $1/ 4\pi^2x^2$,  we have
$$ 
(\delta \Delta^{-1} C_t)(x)=
{1\over 4\pi^2} dx^i\, 
\varepsilon_{ijk} \oint_{l_t} dy^j \int_0^\infty ds\, 
\partial^k\,
{1\over   |\vec x - \vec y (\tau)|^2 + (x^0-y^0(\tau)-s)^2}. 
$$
The integral over $s$ is elementary, and one obtains
\bea
\Gamma(K_s,K_t)&=&\int   K_s \wedge \delta 
\Delta^{-1} C_t =\oint_{l_s}\delta \Delta^{-1} C_t
\label{formula}\\
&=&
{1\over 4\pi^2}\,\varepsilon_{ijk} \oint_{l_s}dx^i\oint_{l_t}dy^j\,
{(x-y)^k\over |\vec x-\vec y|^3}
\left({\pi\over 2}+{\rm arctg}{x^0-y^0\over |\vec x-\vec y|}+
{{x^0-y^0\over |\vec x-\vec y|}\over 
{1 +\left({x^0-y^0\over |\vec x-\vec y|}\right)^2}}\right).
\no
\eea
The main properties of $\Gamma$, needed below, can be deduced from this 
formula. First of all we see that $\Gamma(K_s,K_t)$ is {\it not} 
symmetric in the interchange of  $K_s$ with $K_t$; it has a symmetric
part, represented by the term ${\pi\over2}$ in the integrand, and an
antisymmetric part represented by the $arctg$ and the third term in the
bracket. 
In the symmetric part of $\Gamma$ one recognizes easily the term
${1\over2}\, \#(\vec l_s,\vec l_t)$, where $\#(\vec l_s,\vec l_t)$
indicates  the (integer) linking number 
\footnote{We recall that the linking number of two curves in 
three--dimensional space is given by 
$\#(\vec l_s,\vec l_t)=
{1\over 4\pi}\,\varepsilon_{ijk} \oint_{l_s}dx^i\oint_{l_t}dy^j\,
{(x-y)^k\over |\vec x-\vec y|^3}$.}
of the spatial projections of 
the curves $l_s$ and $l_t$, a crucial feature for what follows. 
This means
that 
\beq
\Gamma(K_s,K_t)={1\over2} \#(\vec l_s,\vec l_t) +\Gamma_{as}(K_s,K_t),
\label{dec}
\eeq
where $\Gamma_{as}$ indicates the antisymmetric part \footnote{The fact 
that the symmetric part of $\Gamma$ is semi--integer 
can also be derived directly from the definition \eref{def}, decomposing
$\Gamma$ in its symmetric and antisymmetric parts, and then using Hodge 
decomposition of the Laplacian and the fact that $\int C_s\wedge C_t$
is integer.}.
This implies in particular that
\beq
\label{sym}
\Gamma(K_s,K_t)+\Gamma(K_t,K_s) \in {\bf Z}. 
\eeq
From \eref{formula}  one sees also that $\Gamma_{as}$ vanishes if $l_s$ and
$l_t$ are at equal times, $x^0(\sigma)=y^0(\tau)= {\rm const.}$, or
if one curve is compact and the other moves to space--like infinity.
In the last case, actually, also $\#(\vec l_s,\vec l_t)$ vanishes.

Finally we note that the property \eref{sym} allows to rewrite the 
imaginary parts of the 
effective actions, apart from integer multiples of $2\pi$, as
\bea
\label{s1}
{\rm Im}\, S_{eff}
&=&-\sum_{s>t}\,(e_sg_t-e_tg_s)\,\Gamma(K_s,K_t)-\sum_s e_sg_s 
\Gamma(K_s,K_s)
\quad  [{\bf Z}_4-{\rm theory}]\\
{\rm Im}\, S_{eff}&=&  -\sum_{s>t}\,(e_sg_t-e_tg_s)\,\Gamma(K_s,K_t)
\quad \phantom{i\sum_s g_s \Gamma(K_s,K_s)} 
[SO(2)-{\rm theory}].
\label{s2}
\eea
Here we used the Dirac quantization condition \eref{2.12}
for the ${\bf Z}_4$--theory and the Schwinger--Zwanziger condition 
\eref{2.37} for the 
$SO(2)$--theory. From these formulae one sees eventually that the unique 
difference in the effective actions of the two theories is represented
by an imaginary part which describes the (diagonal) self--interactions
of the $s$--th dyon with itself, i.e. the second sum in \eref{s1}. It is 
precisely this term which will give rise to spin--statistics 
transmutation for {\it elementary} dyons.

Notice, however, that in the off--diagonal terms, even if formally 
identical,  the coupling constants $(e_sg_t-e_tg_s)$ belong to
$2\pi {\bf Z}$ for the ${\bf Z}_4$--theory, and to $4\pi {\bf Z}$ for the
$SO(2)$--theory. 

\section{Spin of dyon fields}

In this section we outline the derivation of the spin of the $r$--th
dyon species, relying on the construction of 
dyon quantum field operators sketched in the previous section.
The technical details of the derivation, which relies on the properties of
the functional $\Gamma$ displayed above, are relegated to the appendix.

First of all we remark that, except for the rotation symmetric choice of $E$,
the rotation group is not unitarily implementable in 
${\cal H} (E)$. This is due to the fact for a generic rotation ${\cal R}$ 
the behaviour at infinity of the rotated electric 
distribution, $E_{\cal R}$, differs from that of $E$  and, 
as noticed in the previous section, this implies that the 
Hilbert spaces ${\cal H} (E)$ and ${\cal H} (E_{\cal R})$ are orthogonal to 
each other.

However, since a rotation of an integer multiple of $2\pi$ around 
an arbitrary axis
leaves all local observables invariant, in each sector 
${\cal H}_q^r (E)$ it must be represented by a phase, by Schur's lemma. 
Hence, if we denote 
with $U(2\pi)$ the unitary operator which represents a $2\pi$ rotation, one has
\beq
U(2\pi)\, {\cal H}^r_1 (E) = e^{2\pi i s_r(E)}\, {\cal H}_1^r (E), 
\label{4.1}
\eeq
or 
\beq
U(2\pi)\,\hat\phi_r(E^x)\, U^+(2\pi) = e^{2\pi i s_r(E)}\,\hat\phi_r(E^x),
\eeq
where $s_r(E)$ is identified with the spin {\it mod} 1, also called 
spin--type of the dyon field $\hat\phi_r(E^x)$, see \cite{6}. 

According to 
standard arguments one expects $s_r(E)$ to be integer or half--integer.

The correct definition of the action of the operator $U(2\pi)$ requires
some care since the support of
the field $\phi_r (E^x)$ extends to infinity; an improper definition could 
retain erroneously contributions from infinity (see \cite{20} for a
discussion of the related problems).
In dealing with non--local fields a standard procedure is to introduce 
a localized version of $U(2\pi)$, denoted by $U_L(2\pi)$. By definition 
this operator acts trivially 
outside a spatial sphere of radius $L+1$, centered at the origin, it induces a 
$2\pi$--rotation inside a sphere of radius $L$ and it interpolates smoothly 
in between. 
The operator $U(2\pi)$ is then defined as the {\it weak} limit of 
$U_L(2\pi)$ as $L\rightarrow \infty$. 

The simplest correlation function which allows to test the spin of the 
$r$--th dyon is the two--point function. In fact, according to the above 
considerations we can write
\beq
\label{rel}
\lim_{L\rightarrow \infty} \langle \Omega \vert  \bar{\hat\phi_r}(E^y)
\, U_L(2\pi)\,\hat\phi_r(E^x)\,U^+_L(2\pi) \vert \Omega \rangle=
e^{2\pi i s_r(E)}   
\langle \Omega \vert \bar{\hat\phi_r}(E^y)
\hat\phi_r(E^x)\vert \Omega \rangle.
\eeq

Our purpose is to compute $s_r(E)$, using this formula.
For the correlator at the r.h.s. with $x^0 > y^0$ we have already 
a convenient euclidean
path--representation, given in \eref{3.10}; in the correlator at
the l.h.s., for fixed $L$ the field $\hat\phi_r(E^x)$ appears rotated according 
to the above prescription for $U_L$, while the field 
$\bar{\hat\phi_r}(E^y)$ is unchanged. Taking a look at the representation
\eref{3.10} we see that, in the limit $L\rightarrow\infty $, the currents 
${\bf J}_{(s)}$ and $J_{(r)}$ in the integrand are also unchanged, because 
they are of compact support. The unique ingredients which are of 
non--compact support are the currents $\gamma^x$ and 
$\gamma^y$, but, since only $\hat\phi_r(E^x)$ gets rotated, it is 
only the curve $\gamma^x$ which goes over in a curve $\gamma_L^x$.  
For the correlator at the l.h.s. of \eref{rel} we can therefore write the 
following euclidean path--representation $(x^0>y^0)$:
\beq
\label{rot}
N_L(E)
\int {\cal D} \nu_E (\gamma^y)\,{\cal D} \nu_{E_L}(\gamma^x_L)
\int \,{\cal D} \mu (J_{(r)}) \prod_s {\cal D} \mu ({\bf J}_{(s)})\, 
e^{-S_{eff}(j_L)}.
\eeq
$N_L (E)$ is a normalization constant, depending on $L$ and $E$, 
ensuring that in the limit $L\rightarrow \infty$ the action of the $2\pi$ 
rotation reduces to a multiplicative phase factor. The
rotated Mandelstam strings $\gamma_L^x$ are weighted by the measure 
${\cal D} \nu_{E_L} (\gamma^x_L)$, obtained from ${\cal D} \nu_E (\gamma^x)$ 
through a localized $2\pi$ rotation as discussed above.

The effective action depends now on the ``rotated" currents $j_L$ which are 
defined precisely as in \eref{jtot}, with the unique difference that
the curve $\gamma_r$ is replaced by 
\beq
\label{grot}
\gamma_r^L=\gamma^x_L-\gamma^y +J_{(r)}.
\eeq
One can also write 
$$
j_{L\alpha}=j_\alpha +e_{r\alpha}S_L,
$$
where
\beq
S_L \equiv \gamma^x_L-\gamma^x
\label{sl}
\eeq
is a closed curve confined to the region $L \leq|\vec r|\leq L+1$, 
which, as $L\rightarrow \infty$, becomes
infinitely extended and gets placed at infinity.

In comparing the rotated and unrotated correlation functions,
respectively \eref{rot} and \eref{3.10},
one has to evaluate the behaviour of 
$S_{eff}(j_L)-S_{eff}(j)$ as $L \rightarrow \infty$. 
Since, as remarked above, the real part 
of this difference can not give rise to a change of spin--type, 
the spin of the $r$--th dyon species is given by
\beq
s_r=
{1\over 2\pi } \lim_{L\rightarrow \infty} 
 {\rm Im} \left(S_{eff}(j_L)-S_{eff}(j)\right),  
\label{spin}
\eeq
provided this limit is independent of the path--integration variables.
This turns out to be true, as shown in the appendix, where we give also a 
heuristic argument suggesting that one can choose $N_L(E)$ such that
$N_L(E) {\cal D} \nu_{E_L}(\gamma^x_L) exp[-{\rm Re}\,S_{eff}(j_L)]$ approaches
${\cal D} \nu_E(\gamma^x) exp[-{\rm Re}\,S_{eff}(j)]$ as $L \rightarrow \infty$.
The crucial ingredients of the computation of the limit \eref{spin} 
are the linking numbers 
appearing in the imaginary parts of the effective action, and 
an appropriate regularization of the (ultraviolet) divergences showing
up in the $r$--th self--interaction in formula \eref{s1}, which is the 
unique term which eventually gives a non--vanishing contribution.
The final result amounts to the difference of the 
self--linking numbers of two ribbons (see the appendix) and it reads
\bea\label{sz4}
\lim_{L\rightarrow \infty} 
 {\rm Im} \left(S_{eff}(j_L)-S_{eff}(j)\right)&=& 
\,(1/2 \, mod\,\, {\rm{\bf Z}})\,e_rg_r
                               \qquad  [{\bf Z}_4-{\rm theory}],\\
\lim_{L\rightarrow \infty} 
 {\rm Im} \left(S_{eff}(j_L)-S_{eff}(j)\right)&=& 0
\phantom{\,(1/2 \, mod\,\, {\rm{\bf Z}})\,e_rg_r} 
                                      \qquad [SO(2)-{\rm theory}].
\label{sso2}
\eea

So there is no spin--transmutation in the $SO(2)$--theory, while in the 
${\bf Z}_4$--theory
the $r$--th dyon carries spin--(type)
\beq
\label{result}
s_r={e_rg_r\over 4\pi}={n_{rr}\over 2},
\eeq
which is integer or half--integer depending on whether the integer
$n_{rr}$ appearing in the Dirac quantization--condition \eref{2.12} is
even or odd.

\subsection {On the spin addition rule for dyons}

We conclude this section with a remark on the spin addition 
rule.

Let us consider in ${\bf Z}_4$ theories the state $\vert\psi\rangle$ 
corresponding, as explained in section 
2, to the euclidean field $\prod_r \phi_r (E^{x_r}_{(r)})$, with 
the electric distributions $E^{x_r}_{(r)}$ supported in cones with
pairwise disjoint supports. We suppose for simplicity that each species of 
dyons appears once, i.e. $r=1,...,N.$  
If the space--positions $\vec x_r$ are lying all in 
a small region and the time--coordinates practically coincide, then
one might consider $\vert\psi\rangle$ as a state representing
a ``multidyon composite". We investigate here the spin of this 
composite.
  
The computation of the spin-type of $\vert\psi\rangle$ can be performed
with the techniques sketched above. The simplest 
correlation function which allows to test its spin involves 
$\vert\psi\rangle$ and a product of ``compensating" fields carrying 
charges opposite to those of $\vert \psi \rangle$ and 
electric field distributions which ensure that the total flux at space 
infinity decays faster than Coulombic.
The computation of the spin reduces again to 
the evaluation of a limit like \eref{spin}. 
A priori it involves the self--linking numbers of the ribbons
associated to the deformed curves $\vec\gamma_r^ L$ 
defined as above, and the linking numbers 
among all pairs of these curves. 
Assume for simplicity that also the deformed electric 
distributions $E_{(r) L}$ have pairwise disjoint supports. 
In this geometry, 
as $L \rightarrow \infty$, the linking numbers among different curves 
vanish and 
only the self--linking numbers of the ribbons contribute to the spin--type 
of $\vert \psi \rangle$ which, according to \eref{result}, is then given by
\beq
s_{\vert \psi \rangle}={1 \over 4 \pi} \sum_r e_r g_r \,\, mod\,
{\rm{\bf Z}}
\eeq
i.e. we obtain the standard spin addition rule for ${\bf Z}_4$--dyons. 
For $SO(2)$--dyons the spin--type of this composite remains again integer.

On the other hand one might consider in a sense the opposite case of a state
$\vert \tilde\psi \rangle$ corresponding to the composite euclidean field
\beq
\int {\cal D} \nu_E(\gamma^x) \prod_r \phi_r(x) e^{i \ve^{\alpha \beta} 
e_{r \alpha} \int \gamma^x \wedge A_{\beta}},
\eeq
where all fields are located at the same point $x$ and carry a common
Mandelstam--string $\gamma^x$.
The derivation of the spin--type of $\vert \tilde\psi \rangle$ follows
the strategy developed above and one has again to evaluate a 
limit like \eref{spin}. As explained in the appendix each functional 
$\Gamma$ in formulae \eref{s1} and \eref{s2} contributes with a factor
of 1/2 to this limit. This gives for ${\bf Z}_4$--dyons, according to 
\eref{s1},
\beq
\label{comp}
s_{\vert \tilde\psi \rangle}=-{1 \over 4 \pi} \left[\sum_{s>t} 
( e_s g_t- e_t g_s)
+ \sum_s e_s g_s\right] ={1 \over 4 \pi} \left(\sum_s e_s\right) 
\left(\sum_t g_t\right) \,\, mod\, {\rm{\bf Z}},
\eeq
where we used Dirac's quantization condition.
This result coincides with that obtained in quantum--mechanical 
calculations of the spin--type of dyon composites \cite{2,3}.
In general, however, $s_{\vert \psi \rangle} \neq s_{\vert \tilde\psi 
\rangle}$ mod {\bf Z}, although the two states have the same total 
electric and magnetic charges. 

The above results indicate that at the 
QFT--level the spin--type of a 
multi--dyon state does depend not only on its total charges, but also on the 
specific asymptotic behaviour of the soft--photon clouds surrounding the 
dyons \footnote{In the framework of the algebraic approach to relativistic 
QFT this suggests that the spin--type can depend not only on the 
charge--class \cite{26}, but also on the specific superselection sector 
within that class.}.     

For $SO(2)$--dyons \eref{s2} leads instead to 
\beq
s_{\vert \tilde\psi \rangle}=-{1 \over 4 \pi} \sum_{s>t} 
(e_s g_t- e_t g_s) \,\, mod\, {\rm{\bf Z}}.
\label{fff}
\eeq
Here there are no diagonal contributions to spin but, 
a priori, one has now mixed contributions. However, since  
the relevant quantization condition is the Schwinger--Zwanziger condition
we have  $e_sg_t-e_tg_s \in 4\pi {\bf Z}$ and \eref{fff} 
becomes an integer. Therefore, also for these dyon composites there is no 
spin--transmutation in the $SO(2)$--theory.

\section{Statistics of dyon fields}

In this section we perform the 
analysis of the statistics of $r$--th dyon field; more precisely, we
discuss the sign appearing in the commutation relation
\beq
\label{5.1}
\hat\phi_r (E^x) \hat\phi_r (E^{\prime y}) = \pm \hat\phi_r (E^{\prime y}) 
\hat\phi_r (E^x), 
\eeq
which holds provided $x^0=y^0$ and the support of $E^x$ is disjoint 
from the support of $E^{\prime y}$.

Such a condition is never satisfied if we consider fields corresponding to 
the same electric distribution $E$; for this choice of electric 
distributions, in charged sectors of gauge theories with infrared 
QED--like behaviour, 
in \cite{23} it has been proposed to consider only asymptotic commutation 
relations. 

Here we wish to consider the simpler situation described above, where the 
supports of $E$ and $E^\prime$ are given by disjoint cones.
We derive the sign appearing in \eref{5.1} analysing the monodromy
properties of the euclidean correlation functions of dyon fields under 
their exchange.
Four--fields vacuum expectation values of the form
\beq
\label{5.0}
\langle \Omega \vert \bar{\hat\phi_r}(E^z) \bar{\hat\phi_r}(E^{\prime w})
\hat\phi_r(E^x) \hat\phi_r(E^{\prime y}) \vert \Omega \rangle,
\eeq 
with $x^0 \nearrow y^0$, are the simplest correlation functions allowing 
to determine the statistics of dyon fields with euclidean methods as 
follows.
For $z^0<w^0<x^0$ the v.e.v. \eref{5.0} admits a 
representation as expectation value of euclidean fields given by
\bea\nonumber
&&\left\langle \int {\cal D} \nu_E (\gamma^z) \int {\cal D}
\nu_{E^\prime}
(\gamma^w) e^{-i\varepsilon_{\alpha\beta}
\int (e_r^\alpha (\gamma^z +\gamma^w)) \wedge A^\beta}
\bar\phi_r (z) \bar\phi_r (w)\cdot\right.\\
&&\left. \int{\cal D} \nu_{E}
(\gamma^x) \int{\cal D} \nu_{E^\prime} (\gamma^y)
e^{i\varepsilon_{\alpha\beta} \int(e^\alpha_r (\gamma^x +\gamma^y))
\wedge
A^\beta} \phi_r (x) \phi_r (y) \right\rangle
\label{5.2}.
\eea
This expectation value can in turn be rewritten in terms of path--integrals
over currents, which  involve two additive terms corresponding to the two 
admissible contractions 
of the four scalar fields appearing in \eref{5.2}:
\beq
\int {\cal D} \nu_E (\gamma^z){\cal D} \nu_E(\gamma^x)
{\cal D} \nu_{E^\prime}(\gamma^w){\cal D} \nu_{E^\prime}(\gamma^y)
\int {\cal D} \mu (J_{(r)}) {\cal D} \mu (J^\prime_{(r)})
\prod_s {\cal D} \mu ({\bf J}_{(s)})
e^{-S_{eff}(j)} +\{x\leftrightarrow y\}.
\label{four}
\eeq
Following the notations of \eref{jtot}--\eref{gr} we have here the 
insertion of the non--compact curve
\bea
\nonumber
\gamma_r&=&\gamma^x -\gamma^z +J_{(r)}
+\gamma^y -\gamma^w + J^\prime_{(r)}\\
dJ_{(r)}&=&\delta_x -\delta_z\nonumber\\
dJ^\prime_{(r)}&=&\delta_y -\delta_w.
\label{nonso}
\eea
The second term in \eref{four} is obtained from the first one by 
interchanging $x$ and $y$. Denoting with $\widetilde j$ the total current 
obtained through this interchange, this means in particular that the 
exponential in the second term is given by $exp(-S_{eff}(\widetilde j))$.
$\widetilde j$ differs from $j$ only through the insertion of the non--compact
curve, $\widetilde\gamma_r$, which is obtained from $\gamma_r$ with the
replacement $x\leftrightarrow y$.

As in the derivation of the spin we must be careful in handling the 
``behaviour at  infinity" and, according to the
treatment adopted in \cite{21}, we define the exchange of the fields 
$\phi_r(E^x)$ and $\phi_r(E^{\prime y})$ in the above four--point correlation 
function as follows. We introduce a 
deformation of $E^x$ and $E^{\prime y}$ acting trivially outside
a ball of radius $L+1$, exchanging $E^x$ and $E^{\prime y}$ within a ball of 
radius $L$ and interpolating smoothly for intermediate radii. 
We further require that the support of the deformed 
electric distributions, $E^y_L$ and $E^{\prime x}_L$, are still
disjoint for sufficiently large $L$. At the level of Mandelstam strings,
this deformation maps the currents $j$ and $\widetilde j$ appearing in 
\eref{four} into deformed currents $j_L$ and $\widetilde j_L$. In particular,
we have
\bea
\nonumber
\gamma_{rL}&=&\gamma^x_L -\gamma^z +\hat J_{(r)}
+\gamma^y_L -\gamma^w + \hat J^\prime_{(r)}\\
d\hat J_{(r)}&=&\delta_y -\delta_z\nonumber\\
d\hat J^\prime_{(r)}&=&\delta_x -\delta_w,
\nonumber
\eea
and similarly for $\widetilde\gamma_{rL}$. As $L\rightarrow \infty$ we have
\bea
\gamma^x_L&\rightarrow& \gamma^y\no\\
\gamma^y_L&\rightarrow& \gamma^x\no\\
\gamma_{rL}&\rightarrow& \widetilde \gamma_r\no\\
\widetilde\gamma_{rL}&\rightarrow& \gamma_r \no,
\eea
and eventually
\bea
\no
\widetilde j_L&\rightarrow& j\\
j_L&\rightarrow& \widetilde j.\no
\eea

We multiply the correlation functions with the above deformed 
electric distributions and Mandelstam strings by a normalization constant 
$N_L(E,E^\prime)$,
playing a role analogous to $ N_L(E)$ in \eref{rot}, and finally we take 
the limit $L \rightarrow \infty$. The constant $N_L(E,E^\prime)$ should be 
chosen in such a way that as a result of the above operations one obtains 
the original correlation function multiplied by a sign $\pm= e^{i 2 \pi 
\theta},\, \theta=0,1/2$.

From a standard 
argument of the reconstruction theorem one can infer that this is the sign 
appearing in \eref{5.1}, determining the statistics of dyon fields, and 
$\theta$ is their statistics parameter.

The calculations are then similar to those performed to derive the spin. 
As $L\rightarrow \infty$ in the rotated correlator the first (second) term in 
\eref{four} should go over in the second (first) term, apart from the 
overall phase $e^{i 2 \pi \theta}$.
The limits relevant to determine the statistics are then again the ones 
related to the imaginary part of the effective action and the 
calculations, reported in the appendix, give the 
results: 
\bea
\label{stz4}
2 \pi \theta&=&\lim_{L\rightarrow \infty}
 {\rm Im} \left(S_{eff}(j_L)  
-S_{eff}(\widetilde j)\right) \nonumber \\
&=&\lim_{L\rightarrow \infty}
 {\rm Im} \left(S_{eff}(\widetilde j_L)- S_{eff}(j) \right) 
\nonumber \\
&=& \,(1/2 \, mod\,\, {\rm{\bf Z}})\,e_rg_r
                         \qquad  [{\bf Z}_4-{\rm theory}],\no\\
\eea
\bea
\label{stso}
2 \pi \theta&=&\lim_{L\rightarrow \infty}
 {\rm Im} \left(S_{eff}(j_L)  
-S_{eff}(\widetilde j)\right) \nonumber \\
&=&\lim_{L\rightarrow \infty}
 {\rm Im} \left(S_{eff}(\widetilde j_L)- S_{eff}(j) \right) 
\nonumber \\
&=& 0 \phantom{\,(1/2 \, mod\,\, {\rm{\bf Z}})\,e_rg_r}
                                      \qquad [SO(2)-{\rm theory}].
\eea
As shown in the appendix the non--vanishing contribution appearing in 
${\bf Z}_4$--theories comes from
the self--linking number of the ribbons associated to 
(the three--dimensional spatial parts of)
the compact closed currents $j_L-\widetilde j$ and 
$\widetilde j_L-j$. 
These ribbons exhibit an odd number of crossings and therefore their 
self--linking number is odd \cite{kur,22}. (In the simplest case the 
ribbon is associated to a curve whose projection on an arbitrary 
two--dimensional plane has the shape of the symbol ``$\infty$".)

From \eref{stz4} and \eref{stso} one derives that only ${\bf 
Z}_4$--dyon theories with $e_r g_r = 2\pi n_{rr}$ and $ n_{rr}$ odd exhibit 
statistics transmutation.  
The above results are then consistent with the standard 
spin--statistics connection, even if spin--statistics transmutation occurs.

Actually, since the relevant calculation is reduced to a computation of 
self--linking numbers, this connection emerges in a form very similar to 
the ``proof" of the spin--statistics theorem proposed by Wilczek and Zee
\cite{24} for anyons in 
2+1 dimensions (See also \cite{25}  for a weaker analogy with skyrmions).
In fact, these authors associate to every euclidean
worldline of an anyon a line of electric flux and an infinitesimally 
displaced line 
of vorticity; these two lines define a ribbon, and they 
analyse the spin of the anyon in terms of the self--linking number of a 
$2\pi$--twist of the ribbon, and the statistics of the anyon in terms of the 
self--linking number of a two--ribbon exchange. These self--linking
numbers are shown to be identical by a simple geometrical 
argument \cite{22}.

Since in dyon theories we are working in 3+1 dimensions, the matter appears
to be quite different. However, as shown before, choosing the Dirac--strings 
along the time direction  the relevant phase factors arise from a 
projection of the quantum mechanical trajectories of dyons and 
of Mandelstam strings in  a fixed--time 3--space.  
After this projection the calculations involved in 
establishing the spin--statistics connection in 3+1 dimensions  
become very similar to those of the (2+1)--dimensional case.

As noticed in section 4.1, however, the situation is somewhat different for 
what concerns the spin addition rule. This is 
a consequence of the singular infrared behaviour 
of (3+1)--dimensional dyon systems, whose handling requires a cancellation of 
fluxes at infinity.  This feature has no analogue in (2+1)--dimensional 
systems of anyons and (3+1)--dimensional systems of skyrmions.

\section{Appendix}
\subsection{Computation of the dyon spin}

In this subsection we estimate the behaviour of 
$$
\Delta_L\equiv
S_{eff}(j_L)- S_{eff}(j),
$$
in the limit $L \rightarrow \infty$,
relevant for spin--transmutation of the $r$--th dyon, proving formulae
\eref{sz4} and \eref{sso2}.

We begin with the analysis of the real part, which is the same for
$SO(2)$ and $Z_4$--theories. Using \eref{Z} (or \eref{S}) we get 
$$ 
{\rm Re}\,\Delta_L=
e_{r\alpha} S_L \wedge * \Delta^{-1} j_\alpha + 
{1\over 2}
e_{r\alpha}e_{r\alpha} S_L\wedge * 
\Delta^{-1}S_L.  
$$
Heuristically one can argue that the second term can be compensated by the 
constant $N_L (E)$, since 
it depends only on currents weighted by the measures ${\cal D} \nu_{E_L}
(\gamma_L^x)$ and ${\cal D}\nu_E(\gamma^x)$ and that
the first term vanishes in the 
limit $L\rightarrow \infty$, by scale arguments. The reason why one expects
this behaviour is that at 
large scales 
$\gamma_L^x$,  $\gamma^x$ and $\gamma^y$ are weighted by measures peaked 
around $E_L^x$,  $E^x$ and $E^y$, respectively, as follows from 
eq. \eref{3.7}. A mean--field treatment then would give the desired result, 
since $E^x-E^y$ has a dipole--like decay at space--infinity.
That the real part of the effective action can not anyway
contribute to spin--statistics transmutation has already been anticipated
in text.

We turn now to the imaginary part, starting with the $Z_4$--theory.
In the computation which follows we will make repeated use of the 
properties of the functional $\Gamma$ derived in subsection 3.1. 

For ${\rm Im}\,S_{eff}(j)$ we use the formula \eref{s1}. Then 
${\rm Im}\,S_{eff}(j_L)$ is given by the same formula with the unique
difference that $K_r$ is replaced with $K_r +S_L$, with $S_L$ given in
\eref{sl}. Taking advantage of the bilinearity of the functional $\Gamma$
one  gets (a part from an
integer multiple of $2\pi $, which from now on is always understood)
\beq
\label{proto}
-{\rm Im}\,\Delta_L=\sum_{s\neq r}\,(e_rg_s-e_sg_r)\,\Gamma(S_L,K_s) 
         +\,e_rg_r\left[\Gamma(K_r+S_L,K_r+S_L)-\Gamma(K_r,K_r)\right].
\eeq
Since for $s\neq r$ $K_s={\bf J}_{(s)}$ is a compact surface and $S_L$ goes to 
space--like infinity we have 
$$
\lim_{L\rightarrow \infty}\Gamma(S_L,K_s)=0.
$$ 
On the other hand, for the $r$--th current we have 
$K_r={\bf J}_{(r)}+\gamma_r$, where ${\bf J}_{(r)}$ is compact while
$\gamma_r=\gamma^x-\gamma^y+J_{(r)}$ is not. Therefore, for 
$L\rightarrow \infty$, Im $-\Delta_L$ behaves as
\beq
\label{self}
\,e_rg_r\left[\Gamma(\gamma_r+S_L,\gamma_r+S_L)-
\Gamma(\gamma_r,\gamma_r)\right].
\eeq
Now we remained with the currents $\gamma_r$ and 
$\gamma_r+S_L=\gamma^x_L-\gamma^y+J_{(r)}=\gamma_{rL}$ which correspond 
to connected curves; this means that both functionals
$\Gamma$ appearing in \eref{self} need a proper regularization due to the 
3--space intersection points in the integral \eref{formula}, occurring for
$\sigma=\tau$ \footnote{This regularization is, actually, needed from
the beginning for the functional $\Gamma(K_r,K_r)$, evaluated for two 
identical currents. However, this regularization does not affect the 
vanishing of the terms discussed so far.}.
A standard regularization for such coincident curves is given by a 
framing procedure, in which the
curve appearing say in the first argument of $\Gamma$ gets displaced 
by an infinitesimal 3--space vector $\vec \varepsilon$, 
orthogonal to the curve. Accordingly we replace \eref{self} by the 
well defined expression
\bea
& &\,e_rg_r\left[\Gamma(\gamma_r^\ve+S_L^\ve,\gamma_r+S_L)-
\Gamma(\gamma_r^\ve,\gamma_r)\right]\no\\
&=&\,e_rg_r\left[\Gamma(\gamma_r^\ve,S_L)+\Gamma(S_L^\ve,\gamma_r)
+\Gamma(S^\ve_L,S_L)\right]\no\\
&=&{1\over2}\,e_rg_r 
\left[
\#(\vec\gamma_r^\ve,\vec S_L)+\#(\vec S_L^\ve,\vec\gamma_r)
+\#(\vec S^\ve_L,\vec S_L)\right]\no\\
&=&{1\over2}\,e_rg_r \left(
\#(\vec \gamma_{rL}^\ve,\vec \gamma_{rL})-
\#(\vec \gamma_{r}^\ve,\vec \gamma_{r})\right).
\label{diff}
\eea
With the subscript $\ve$ we indicate the framed currents. In 
the second line we use bilinearity. To obtain the third line we use
the fact that the 
antisymmetric parts of the functionals $\Gamma$ cancel:
$\Gamma_{as}(S^\ve_L,S_L)$ vanishes because the curves $S^\ve_L$ and
$S_L$ stay at equal (constant) time $x^0$; for what concerns 
$\Gamma_{as}(\gamma_r^\ve,S_L)$ and $\Gamma_{as}(S_L^\ve,\gamma_r)$
we observe that the space--intersection points between  $\gamma_r$
and $S_L$ lie along $\gamma^x$, a curve which is at constant time $x^0$.
But since also $S_L$ stays at time $x^0$, the intersection points
do not contribute to the two $\Gamma_{as}$ in consideration. This means 
that the regularization can be removed and one has then trivially
$\Gamma_{as}(\gamma_r,S_L)+\Gamma_{as}(S_L,\gamma_r)=0$. 
In conclusion, in the third line above only the (regularized) linking 
numbers between the spatial parts of the corresponding currents
 survive. The fourth line follows from bilinearity of the
linking number between two curves in three dimensions.

The pair of curves $\vec \gamma_{r}$ and its framed version 
$\vec\gamma_{r}^\ve$ define what is called a ribbon, and the
integer number $\#(\vec \gamma_{r}^\ve,\vec \gamma_{r})$ defines then
the self--linking number of this ribbon, which is a topological invariant
(see e.g. \cite{22} and references therein). So what we are computing
in \eref{diff} is the difference between the self--linking numbers
of the ribbons $(\vec\gamma_{rL},\gamma_{rL}^\ve)$ and 
$(\vec\gamma_{r},\gamma_{r}^\ve)$. From the geometry of the curves
involved it is clear
that, as $L\rightarrow \infty$, this difference is {\it odd}. This leads to
\beq
\label{app}
\lim_{L\rightarrow\infty}{\rm Im}\,\Delta_L=
\,(1/2 \, mod\,\, {\rm{\bf Z}})\,e_rg_r,
\eeq
as stated in the text. 

From this calculation it is also clear that in the $SO(2)$--theory 
we have $\lim_{L\rightarrow\infty}{\rm Im}\,\Delta_L=0$. This is due to the 
fact that in the corresponding effective action
\eref{s2} the self--interaction terms $\Gamma(K_r,K_r)$, which 
eventually led to the non--vanishing result in \eref{app}, are absent. 

More generally we conclude that 
the functional $\Gamma(K_s,K_t)$ can
contribute to ${\rm Im}\,\Delta_L$ as $L \rightarrow \infty$ 
with an additive term of 1/2, and 
therefore to spin--type, only if both currents are non compact, 
corresponding to the insertion of charged fields of the species
$s$ and $t$ in the correlator, and if one rotates say one charged field
of the $s$--type and one of the $t$--type. The case considered above
corresponds to $s=t=r$. 

\subsection{Computation of the dyon statistics}

Here we evaluate the limit 
$$
\lim_{L\rightarrow \infty}
 {\rm Im} \left(S_{eff}(j_L)  
-S_{eff}(\widetilde j)\right),
$$
where the currents are specified in section five, according to the formulae
\eref{jtot}--\eref{gr}. We begin with the ${\bf Z}_4$--theory.
It is convenient to define the closed compact current 
$$
S_L=\gamma_{rL}-\widetilde\gamma_r,
$$
which for $L\rightarrow \infty$ gets placed at infinity, 
because then for 
$ {\rm Im} \left(S_{eff}(j_L)-S_{eff}(\widetilde j)\right)$ we can write
an expression which is formally identical to \eref{proto}. With precisely 
the same steps as above, in particular with the same regularization 
procedure and using the fact that the curves $\gamma^x$, $\gamma^y$ and
$S_L$ stay at the same fixed time $x^0=y^0$, we arrive at the formula
(see the third line in \eref{diff})
\beq
-\lim_{L\rightarrow \infty}
 {\rm Im} \left(S_{eff}(j_L)  
-S_{eff}(\widetilde j)\right)={1\over 2}\,e_rg_r\lim_{L\rightarrow \infty}
\left[
\#(\vec\gamma_r^\ve,\vec S_L)+\#(\vec S_L^\ve,\vec\gamma_r)
+\#(\vec S^\ve_L,\vec S_L)\right].
\eeq
This time $\gamma_r$ is the union of {\it two} connected curves, see 
\eref{nonso}, and the geometry is such that for large enough $L$ the sum
of liking numbers 
$\#(\vec\gamma_r^\ve,\vec S_L)+\#(\vec S_L^\ve,\vec\gamma_r)$ becomes
{\it even}. On the other hand, the self--linking number of the curve
$\vec S_L$, i.e. the linking number of the ribbon $(\vec S^\ve_L,\vec S_L)$,
is {\it odd}. This is due to the fact that the projection of $\vec S_L$
on an arbitrary two--plane exhibits an odd number of crossings \cite{22}. 
In the simplest case this projection corresponds to an ``eight".
This leads to the result
$$
\lim_{L\rightarrow \infty}
 {\rm Im} \left(S_{eff}(j_L)  
-S_{eff}(\widetilde j)\right)=
\,(1/2 \, mod\,\, {\rm{\bf Z}})\,e_rg_r,
$$
quoted in the text. 

In the SO(2)--theory the corresponding limit is 
zero for the reasons quoted above, i.e. the absence of 
self--interactions.

\bigskip  
\paragraph{Acknowledgements.}

This work was supported by the 
European Commission RTN programme HPRN-CT2000-00131.

%%%%%%%%%%%%%%%%%%%%%%%%%%%%%%%%%%%%%%%%%%%%%%%%%%%%%%%%%%%%%%%%%%%%%%%

%\newpage

\end{document}